\documentclass[a4paper,11pt]{article}
\usepackage{aaskaiid}
\usepackage{orcidlink}
\usepackage{multicol}
\usepackage{wrapfig}
\usepackage[leftcaption]{sidecap}
\newcommand\footnoteref[1]{\protected@xdef\@thefnmark{\ref{#1}}\@footnotemark}

\newcommand{\Lbol} {$L_{\rm bol}$}
\newcommand{\ionfrac} {$\chi_{\mathrm{e}}$}
\newcommand{\massloss} {$\dot{M}_{\rm loss}$}
\newcommand{\massacc} {$\dot{M}_{\rm acc}$}
\newcommand{\nEL} {$n_{\rm e}$}
\newcommand{\Mion} {$M_{\rm ion}$}
\newcommand{\hi}{\mbox{H\textsc{i}}}
\newcommand{\hii}{\mbox{H\textsc{ii}}}

\title{Jets and outflows in young stellar objects with the SKAO} %\footnote{Useful Links: (1)SKA sensitivity calculator: \url{https://sensitivity-calculator.skao.int/mid}; (2) Tools and documentation: \url{https://www.skao.int/en/science-users/ska-tools/493/ska-sensitivity-calculators}; (3) SKAO demo of tools: \url{https://drive.google.com/drive/folders/1StTpO_gen33aSgHgNovRpvoHgO2D-d0y?usp=share_link}}}
\ShortTitle{Jets \& Outflows in YSOs with the SKAO}

\author[1]{Sabatini G.\orcidlink{0000-0002-6428-9806}}
\ShortName{Sabatini G. et al.} % shortened name list for header 
\author[2,3]{Busquet G.\orcidlink{0000-0002-2189-6278}}
\author[4]{Carrasco-González C.\orcidlink{0000-0003-2862-5363}}
\author[5]{Rodríguez-Kamenetzky A.\orcidlink{0000-0002-4731-4934}}
\author[1]{Codella C.\orcidlink{0000-0003-1514-3074}}
\author[1]{Podio L.\orcidlink{0000-0003-2733-5372}}
\author[6]{Martínez-Henares A.\orcidlink{0000-0001-5191-2075}}
\author[7]{Girart J. M.\orcidlink{0000-0002-3829-5591}}
\author[8,1]{De Simone M.\orcidlink{0000-0001-5659-0140}}
\author[9]{Cacciapuoti L.\orcidlink{0000-0001-8266-0894}}
\author[10]{Anglada G.\orcidlink{0000-0002-7506-5429}}
\author[11]{Tychoniec L.\orcidlink{0000-0002-9470-2358}}
\author[1]{Giani L.\orcidlink{0000-0002-7327-9132}}
\author[12]{Puravankara M.\orcidlink{0000-0002-3530-304X}}
\author[1]{Bacciotti F.\orcidlink{0000-0001-5776-9476}}
\author[13]{Bachiller R.\orcidlink{0000-0002-5331-5386}}
\author[1,14]{Bianchi E.\orcidlink{0000-0001-9249-7082}}
\author[10]{Bl{\'a}zquez-Calero, G.\orcidlink{0000-0002-7078-2373}}
\author[15]{Bourke T. L.\orcidlink{0000-0001-7491-0048}}
\author[16, 17, 1, 18]{Bovino S.\orcidlink{0000-0003-2814-6688}}
\author[19]{Caselli P.\orcidlink{0000-0003-1481-7911}}
\author[20]{Cavallaro F.\orcidlink{0000-0003-1856-6806}}
\author[14]{Ceccarelli C.\orcidlink{0000-0001-9664-6292}}
\author[2,7]{Diaz-Marquez E.\orcidlink{0009-0005-4546-1104}}
\author[21]{Facchini S.\orcidlink{0000-0003-4689-2684}}
\author[22]{Garufi A.\orcidlink{0000-0002-4266-0643}}
\author[14]{Guidi G.\orcidlink{0000-0002-7002-8928}}
\author[23,24]{Hirota T.\orcidlink{0000-0003-1659-095X}}
\author[25]{Ilee J.~D.\orcidlink{0000-0003-1008-1142}}
\author[20]{Ingallinera A.\orcidlink{0000-0002-3137-473X}}
\author[6]{Jim\'enez-Serra I.\orcidlink{0000-0003-4493-8714}}
\author[19]{Lattanzi V.\orcidlink{0000-0001-9819-1658}}
\author[26]{Lee C.-F.\orcidlink{0000-0002-3024-5864}}
\author[1]{Lippi M.\orcidlink{0000-0001-9185-878X}}
\author[27]{Lupi A.\orcidlink{0000-0001-6106-7821}}
\author[28,29]{Majumdar L.\orcidlink{0000-0001-7031-8039}}
\author[30]{Narang M.\orcidlink{0000-0002-0554-1151}}
\author[10]{Osorio M.\orcidlink{0000-0001-9276-139X}}
\author[1]{Padovani M.\orcidlink{0000-0003-2303-0096}}
\author[19]{Pineda J.\orcidlink{0000-0002-3972-1978}}
\author[22]{Radley I.\orcidlink{0009-0007-2837-8207}}
\author[31]{Riaz B.\orcidlink{0000-0003-3863-4052}}
\author[4]{Rodríguez L. F.\orcidlink{0000-0003-2737-5681}}
\author[32]{S\'anchez-Monge \'A.\orcidlink{0000-0002-3078-9482}}
\author[33]{Sanna A.\orcidlink{0000-0001-7960-4912}}
\author[19]{Spezzano S.\orcidlink{0000-0002-6787-5245}}
\author[34]{Testi L.\orcidlink{0000-0003-1859-3070}}
\author[35,1]{Toci C.\orcidlink{0000-0002-6958-4986}}
\author[36]{Traficante A.\orcidlink{0000-0003-1665-6402}}
\author[12]{Tyagi H.\orcidlink{0000-0002-9497-8856}}
\author[20]{Umana G. M.\orcidlink{0000-0002-6972-8388}}

\affiliation[1]{INAF, Osservatorio Astrofisico di Arcetri, Largo E. Fermi 5, I-50125, Firenze, Italy}
\emailAdd{giovanni.sabatini@inaf.it}
\affiliation[2]{Departament de Física Quàntica i Astrofísica, Universitat de Barcelona, Martí i Franquès 1, 08028 Barcelona,
Catalonia, Spain}
\affiliation[3]{Institut de Ci\'{e}ncies del Cosmos, Universitat de Barcelona, c. Mart\`{i} i Franqu\'{e}s, 1, 08028 Barcelona, Spain}
\affiliation[4]{Instituto de Radioastronomía y Astrofísica, A. Postal 3-72 (Xangari), 58089 Morelia, Michoacán, Mexico}
\affiliation[5]{Instituto de Astronomía Téorica y Experimental (IATE, CONICET-UNC), Córdoba, Argentina}
\affiliation[6]{Centro de Astrobiologia (CAB), CSIC-INTA, Carretera de Ajalvir km 4, Torrejon de Ardoz, 28850, Madrid, Spain}
\affiliation[7]{Institut de Ciències de I’Espai (ICE-CSIC), Campus UAB, Carrer de Can Magrans s/n, 08193 Cerdanyola del Vallès, Catalonia, Spain}
\affiliation[8]{European Southern Observatory, Karl-Schwarzschild-Strasse 2 D85748 Garching bei Munchen, Germany}
\affiliation[9]{European Southern Observatory, Alonso de Cordova 3107, Vitacura, Santiago, Chile}
\affiliation[11]{Leiden Observatory, Leiden University, P.O. Box 9513, 2300RA Leiden, The Netherlands}
\affiliation[12]{Tata Inst. of Fundamental Research, Homi Bhabha Road, Colaba, Mumbai, 400005, India}
\affiliation[13]{Observatorio Astronómico Nacional (OAN-IGN), Alfonso XII 3, 28014 Madrid, Spain}
\affiliation[14]{Univ. Grenoble Alpes, CNRS, IPAG, F-38000 Grenoble, France}
\affiliation[15]{SKA Observatory, Jodrell Bank, Lower Withington, Macclesfield, SK11 9FT, UK}
\affiliation[16]{Department of Chemistry, Sapienza University of Rome, P.le Aldo Moro 5, 00185 Rome, Italy}
\affiliation[17]{Departamento de Astronomía, Facultad Ciencias Físicas y Matemáticas, Universidad de Concepción Av. Esteban Iturra s/n Barrio
Universitario, Casilla 160, Concepción, Chile}
\affiliation[18]{Institute of Structure of Matter (ISM-CNR), Consiglio Nazionale delle Ricerche, 34149 Basovizza, Italy}
\affiliation[19]{Center for Astrochemical Studies, Max Planck Institute for Extraterrestrial Physics, Gießenbachstraße 1, 85748 Garching,
Germany}
\affiliation[20]{INAF, Osservatorio Astrofisico di Catania, Via Santa Sofia 78, 95123 Catania, Italy}
\affiliation[21]{Dipartimento di Fisica, Univ. degli Studi di Milano, Via Celoria 16, I-20133 Milano, Italy}
\affiliation[22]{INAF, Istituto di Radioastronomia, Via Gobetti 101, I-40129, Bologna, Italy}
\affiliation[23]{Mizusawa VLBI Observatory, NAOJ, 2-12 Hoshigaoka-cho, Mizusawa, Oshu-shi, Iwate 023-0861, Japan}
\affiliation[24]{SOKENDAI (The Graduate University for Advanced Studies), 2-21-1
Osawa, Mitaka-shi, Tokyo 181-8588, Japan}
\affiliation[25]{School of Physics and Astronomy, University of Leeds, Leeds, LS2 9JT, UK}
\affiliation[26]{Institute of Astronomy and Astrophysics, Academia Sinica, No. 1-4, Roosevelt Rd., Taipei 106216, Taiwan}
\affiliation[27]{DiSAT, Univ. degli Studi dell’Insubria, via Valleggio 11, 22100 Como, Italy}
\affiliation[28]{National Institute of Science Education and Research, Jatni 752050, Odisha, India}
\affiliation[29]{Homi Bhabha National Institute, Training School Complex, Anushaktinagar, Mumbai 400094, India}
\affiliation[30]{JPL, California Institute of Technology, 4800 Oak Grove Drive, Pasadena, CA 91109, USA}
\affiliation[31]{Universitäts-Sternwarte München, Ludwig Maximilians Univ., Scheinerstra$\beta$e 1, 81679 München, Germany}
\affiliation[32]{Institut de Ciències de l’Espai, CSIC, Campus UAB, Carrer de Can Magrans s/n, 08193 Bellaterra, Barcelona, Spain}
\affiliation[33]{INAF, Osservatorio Astronomico di Cagliari, via della Scienza 5, 09047 Selargius, Italy}
\affiliation[34]{Dipartimento di Fisica e Astronomia ``Augusto Righi'', Univ. di Bologna, Via Gobetti 93/2, 40129 Bologna, Italy}
\affiliation[35]{Departamento de Fisica aplicada III, ETSI Universidad de Sevilla,  Camino de los Descubrimientos, 41092 Sevilla}
\affiliation[36]{INAF, Istituto di Astrofisica e Planetologia Spaziale, Via Fosso del Cavaliere 100, 00133 Roma, Italy}

\abstract{Jets and outflows are ubiquitous phenomena associated with the formation of young stellar objects (YSOs). They play a crucial role in removing angular momentum from the accreting system and in regulating star-formation efficiency. Theoretical studies and observations with ALMA and VLA have shown that jets and winds may have a crucial role in promoting dust growth in the envelope-disc system and in shaping the physical and chemical properties of the surrounding environment. Despite these significant advances, many fundamental questions remain unanswered regarding the acceleration, collimation, and chemical impact of jets and outflows from YSOs. The SKA-project will overcome the limitations of current mm/cm-facilities by enabling high-angular resolution and high-sensitivity cm-observations, crucial for probing jets/outflows near YSOs. Radio recombination lines, combined with proper motions, offer a unique opportunity to study the 3D-kinematics of jets. Non-thermal linearly polarised synchrotron emission will allow measuring magnetic field strength and morphology at unprecedented scales of a few au. Observations of dust emission in outflow cavities will allow studying how dust grows and is eventually transported from the disc to the envelope and back. Finally, the SKA-project will allow exploring the dust composition and chemical enrichment in shocks, where sputtering/shattering of grains cause the release of their mantles and refractory cores in the gas-phase. Complementary to ALMA’s detection of simple and complex organic molecules, the SKAO will probe, for the first time, long carbon chains/rings, several Cl-, Al-, Mg-, and other metal-bearing species (missed by current sub-mm facilities).}

\begin{document}
\maketitle
\newcommand{\actaa}{Acta Astron.} % Acta Astronomica
\newcommand{\araa}{ARA\&A} % Annual Review of Astron and Astrophys
\newcommand{\aar}{A\&ARv} % Astrononmy \& Astrophysics Review
\newcommand{\aapr}{A\&ARv} % Astronomy\&Astrophysics Reviews
\newcommand{\ab}{Astrobiol.} % Astrobiology
\newcommand{\aj}{AJ} % Astronomical Journal
\newcommand{\apj}{ApJ} % Astrophysical Journal
\newcommand{\apjl}{ApJL} % Astrophysical Journal, Letters
\newcommand{\apjs}{ApJSS} % Astrophysical Journal, Supplement
\newcommand{\ao}{Appl. Opt.} % Applied Optics
\newcommand{\apss}{Astro. \& Space Sci.} % Astrophysics and Space Science
\newcommand{\aap}{A\&A} % Astronomy and Astrophysics
\newcommand{\aaps}{A\&AS.} % Astronomy and Astrophysics, Supplement
\newcommand{\baas}{Bull. Am. Astron. Soc.} % Bulletin of the AAS
\newcommand{\caa}{Chinese A\&A} % Chinese Astronomy and Astrophysics
\newcommand{\cjaa}{Chinese J. A\&A} % Chinese Journal of Astronomy and Astrophysics
\newcommand{\cqg}{Class. Quantum Gravity} % Classical and Quantum Gravity
\newcommand{\gal}{Galaxies} % Galaxies
\newcommand{\gca}{Geo. Cosmo. Acta} % Geochimica Cosmochimica Acta
\newcommand{\icarus}{Icarus} % Icarus
\newcommand{\jcap}{JCAP} % Journal of Cosmology and Astroparticle Physics
\newcommand{\jgr}{J. Geophys. Res.} % Journal of Geophysics Research
\newcommand{\jgrp}{J. Geophys. Res. Planets} % Journal of Geophysics Research: Planets
\newcommand{\jqsrt}{J. Quant. Spectrosc. Radiat. Transf.} % Journal of Quantitiative Spectroscopy and Radiative Transfer
\newcommand{\memsai}{Mem. SAIt} % Mem. Societa Astronomica Italiana
\newcommand{\mnras}{MNRAS} % Monthly Notices of the RAS
\newcommand{\nat}{Nature} % Nature
\newcommand{\nastro}{Nat. Astron.} % Nature Astronomy
\newcommand{\ncomms}{Nat. Commun.} % Nature Communications
\newcommand{\nphys}{Nat. Phys.} % Nature Physics
\newcommand{\na}{New Astron.} % New Astronomy
\newcommand{\nar}{New Astron. Rev.} % New Astronomy Review
\newcommand{\physrep}{Phys. Rep.} % Physics Reports
\newcommand{\pra}{Phys. Rev. A} % Physical Review A: General Physics
\newcommand{\prb}{Phys. Rev. B} % Physical Review B: Solid State
\newcommand{\prc}{Phys. Rev. C} % Physical Review C
\newcommand{\prd}{Phys. Rev. D} % Physical Review D
\newcommand{\pre}{Phys. Rev. E} % Physical Review E
\newcommand{\prx}{Phys. Rev. X} % Physical Review X
\newcommand{\prl}{Phys. Rev. Let.} % Physical Review Letters
\newcommand{\psj}{Planet. Sci. J.} % Planetary Science Journal
\newcommand{\planss}{Planet. Space Sci.} % Planetary Space Science
\newcommand{\pnas}{Proc. Natl Acad. Sci. USA} % Proceedings of the US National Academy of Sciences
\newcommand{\procspie}{Proc. SPIE} % Proceedings of the SPIE
\newcommand{\pasa}{PASA} % Publications of the Astron.  Soc. of Australia
\newcommand{\pasj}{PASJ} % Publications of the Astron.  Soc. of Japan 
\newcommand{\pasp}{PASP} % Publications of the Astron.  Soc. of the Pacific
\newcommand{\rmxaa}{RMXAA} % Revista Mexicana de Astronomia y Astrofisica
\newcommand{\sci}{Science} % Science
\newcommand{\sciadv}{Sci. Adv.} % Science Advances
\newcommand{\solphys}{Sol. Phys.} % Solar Physics
\newcommand{\sovast}{Soviet Ast.} % Soviet Astronomy
\newcommand{\ssr}{Space Sci. Rev.} % Space Science Reviews
\newcommand{\aspcs}{Astronomical Society of the Pacific Conference Series} % Universe\newcommand{\uni}{Universe} % Universe

\section{Introduction}\label{sec:intro}
\begin{figure}
\centering
\includegraphics[width=0.95\textwidth]{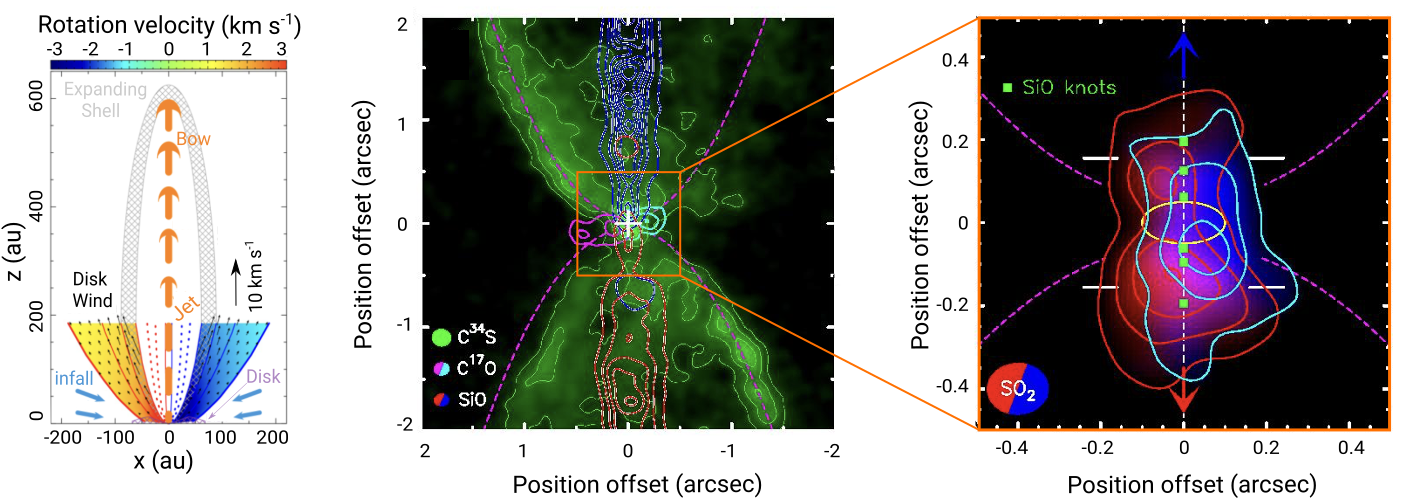}
\vspace{-5pt}
\caption{Left: Sketch of the extended disc wind and jet from a protostar. The wind launches from 4–40 au, while the jet originates at radii $<$0.20~au. The shell (gray cross-hatched) spans from the wind’s inner region to the jet-driven bow shock (adapted from \citealt{Lee21}). Middle Panel: The HH~212 protostellar disc/jet system viewed by ALMA, revealing: (i) the SiO jet (blue and red contours), (ii) the C$^{17}$O rotating envelope (pink and turquoise), and (iii) the cavity walls in C$^{34}$S (green). The white cross indicates HH~212-mm protostar. Right: Zoom-in on blue and red SO$_2$ shows a slow outflow, proposed to represent the disc wind.} 
  \label{fig:diskwinds}
\end{figure}

In the early stages of star formation, the accretion of mass is typically linked to the expulsion of material. This process occurs in young stellar objects (YSOs) across the full range of protostellar masses (from brown dwarfs to massive objects; e.g. \citealt{Frank2014,Anglada2015, Anglada2018}).\\
In Solar-like protostars, for accretion to occur, the angular momentum is removed through ejection from a rotating/accreting disc \citep{Shu1987}, mainly through collimated, and high-velocity ($\sim$100~km~s$^{-1}$) jets. The exact origin of the jet remains elusive, whether it emerges from the innermost part of the disc, through an X-wind \citep{Shang2007} or a dust-free magneto-hydrodynamic (MHD) disc wind 
\citep{Tabone2020}. The high-spatial resolution that can be reached by observing close-by low-mass star-forming regions allows researchers to capture detailed views of jet and disc systems. As an example, Figure \ref{fig:diskwinds} shows how observations of different molecular species lead to an instructive picture of a pristine jet/disc system such as HH~212-mm (e.g. \citealt{Lee21}). Collimated jets propagate along the symmetry axis of the system reaching high velocities up to several hundreds of km~s$^{-1}$. They are surrounded by denser, slower ($\sim$10~km~s$^{-1}$) outflows of a wider opening angle, which can extend over large distances ($\sim$ 0.1 pc), and are commonly traced through low-J CO rotational lines \citep{Lada1985, Lopez-Vazquez24}. Moving to the high-mass regimes ($M_\star\geq$8~M$_{\rm \odot}$), the picture is less detailed due
to the observational limits introduced by the large distances of the high-mass star forming regions (distances $d\gtrsim$1~kpc; e.g. \citealt{Sanhueza19, Morii23, Urquhart22, Molinari25}). High-mass (proto)stars continue to accrete material even after reaching the main sequence and initiating hydrogen fusion, and they predominantly form within stellar clusters \citep[e.g][and references therein]{Motte2018,Beuther2025}, with formation timescales shorter than those of Sun-like YSOs \citep[e.g.][]{Gerner14, Gieser21, Sabatini21}.
In both high- and low-mass regimes, the ejected material interacts with the dense surrounding interstellar medium (ISM). The resulting interaction drives high-velocity ($\geq$ 10 km s$^{-1}$) shocks, which subsequently sputter and vaporise the grain cores and mantles \citep[e.g.][]{Caselli1997, Jimenez-Serra08}. This leads to a significant increase in the abundance of several molecular species, as evidenced by surveys at (sub-)millimeter wavelengths \citep[e.g.][]{Bachiller2001,Codella2010}. Shocks along protostellar jets are unique interstellar laboratories, where species usually locked into dust refractory cores are desorbed and can be investigated in gas-phase. Shocks are also opportunities to explore the chemical processes occurring in environments where the gas is both dense and warm and where its composition has been significantly enriched by the by-products of dust \citep[e.g.][]{Ceccarelli2023}.\\
This Chapter provides an overview of the physical and chemical properties of jets and outflows in YSOs (Sect.~\ref{sec:intro}), spanning wavelengths from the optical to the cm regimes across the full range of stellar masses. The subsequent Sections then detail how the Square Kilometre Array Observatory (SKAO) is positioned to play a pivotal role in resolving the outstanding questions that remain open in this field.

\subsection{Previous results in the optical, IR, (sub-)mm \& radio domains} \label{sec:previous}
\subsubsection{IR/optical facilities} \label{sec:IR-telescopes}
The infrared and optical spectrum provides insight into the warm and hot component ($>$ 500 K) of the outflows. Notably, the most abundant molecule, H$_2$, is exclusively observed at $\lambda< 30~\mu$m. Ground-based optical and IR studies of YSOs are often limited to more evolved Class II systems \citep[e.g.][]{Frank2014}, since the envelope dissipation enables studies of their jets close to the launching zone. Studies of Class I/II jets with the {\it Very Large Telescope} (VLT) revealed depleted elemental abundances, suggesting a significant presence of dust. The [Fe II] lines, in particular, allowed accurate estimates of shock conditions and elemental abundances \citep{Podio2006,Podio.Eisloeffel.ea2011, Giannini.Antoniucci.ea2015}. Velocity-resolved studies have been crucial for revealing the structure of jets and the presence of wide-angle winds \citep[e.g.][]{Pontoppidan.Blake.ea2011, Birney.Dougados.ea2024}. 
{\it Hubble Space Telescope} (HST) high-angular resolution capabilities allowed the measurement of jet widths and rotation close to their launching zones, providing key insights into their collimation and launching mechanism \citep{Bacciotti.Ray.ea2002, Erkal.Nisini.ea2021}.\\
Spectroscopy with the  {\it Infared Space Observatory} (ISO) and {\it Spitzer} Space Telescope enabled systematic access to H$_2$ rotational transitions in Class 0/I systems, allowing the estimation of mass and physical conditions of the outflowing gas \citep[e.g.][]{vanDishoeck2004a, Maret.Bergin.ea2009}. With the advent of the {\it Herschel} Space Observatory, access to the far-IR range has been opened, where the bulk of the outflow cooling is taking place. Using the emission lines of CO and H$_2$O rotational transitions, it was possible to obtain a quantitative and qualitative picture of the outer layers of the outflow streams. A scenario of chemical evolution of the outflow could be modelled, from predominantly molecular to atomic and ionised \citep{Podio2012,Nisini.Santangelo.ea2015}, with two clear temperature components of CO gas, and the presence of hot water in shocks close to the protostars \citep[e.g.][]{Kristensen2012, Manoj13, Watson.Calvet.ea2016, Karska.Kaufman.ea2018}.
SOFIA airborne mission delivered velocity-resolved information on the jets at far-IR \citep{Lefloch.Gusdorf.ea2015}. 
A breakthrough in studies of the protostars and their jets is being made by {\it James Webb} Space Telescope (JWST), which enables observations of even the most embedded systems at sub-arcsecond scales in near and mid-IR. The high sensitivity of JWST enabled the detection of ionised jets even in the very low-mass protostars \citep{Narang24}, at the same time showing purely molecular flow in other systems \citep{Ray2023}, suggesting a more complex picture of the youngest outflows than just a time-related chemical evolution. Another surprise provided by JWST was a detection of H recombination lines \citep{Federman24} and dust continuum emission in the jet \citep{Caratti2024} likely associated with dust lifted from the disc. Precise characterisation of the jets in Class 0/I systems, linking their accretion and ejection properties, is now possible \citep{Nisini2007, Tychoniec24}. 
The upcoming Extremely Large Telescope (ELT) will deliver ultra-high spatial and spectral resolution at optical/NIR wavelengths, providing direct access to the innermost ($<$1~au) launch regions of the disc wind via overtone CO bands and other diagnostics, resolving jet rotation and collimation on sub-au scales.
In summary, infrared studies of jets and outflows present mounting evidence for extended protostellar disc winds \citep{Podio.Eisloeffel.ea2011, Tychoniec24, Delabrosse.Dougados.ea2024}. At the same time, the temperature structure of outflows can be
identified by access to the main cooling agents like CO, H$_2$O, and H$_2$ \citep[e.g.][]{Maret.Bergin.ea2009, Goicoechea.Cernicharo.ea2012, Manoj16}. 

\subsubsection{(Sub)-millimetre telescopes} \label{sec:mm-telescopes}
Low-J transitions of CO in the (sub-)millimetre are widely used to study the morphology and kinematics of outflows \citep[e.g.,][]{Lada1985,Arce2006,Dunham2014,Feddersen2020, deValon22, Bacciotti25, Blazquez-Calero26}. Classical large-scale bipolar molecular outflows, mainly composed of swept-up ambient material, travel at velocities of $\sim$10 km~s$^{-1}$, have sizes from $\sim$0.01 to a few parsecs, masses of $\sim$0.1 to 1 $M_\odot$, and dynamical timescales of 10$^3$–10$^6$~yr \citep{Lada1985,Cabrit1997,Bally2007, Arce2007, Bally2016}. In addition, low-J molecular transitions at (sub)mm-wavelengths have revealed a highly-collimated extremely high velocity (EHV; $\sim$100 km~s$^{-1}$) molecular component of the outflows in the youngest protostars, also referred to as (cold) molecular jets \citep{Bachiller1990,Lee2020}. Highly sensitive, very high angular resolution observations of this cold component, such as those now possible with ALMA, allow us to map intermediate velocities, bridging the gap between the velocities of the EHV jets and the slower velocities typical of the entrained molecular material. This type of study can provide crucial information about how the jet interacts with its surrounding medium \citep{Blazquez-Calero26}. Another new advance enabled by modern millimetre interferometers, particularly ALMA, is the discovery of small-scale ($<$2000 au) molecular flows \citep{Pascucci2023} that rotate in the same sense as the disk and appear to be launched from radii of $\lesssim$100 au \citep{Launhardt2009, Bjerkeli2016, Zapata2015, deValon22, Lee2018, Tabone2017, Zhang2018, Lopez-Vazquez23, Lopez-Vazquez24}, up to the case study of HH~212 where \citet{LeeCF17} imaged a rotating SiO jet at the remarkable resolution of 10~au.\\
In addition, some outflows, identified as \textit{chemically active}, display strong enhancements of additional molecular species. These enhancements are driven by endothermic chemical reactions and sputtering (gas-grain collisions) and shattering (grain-grain collision) of grain mantles and cores that inject atoms and molecules into the gas phase \citep[e.g.][]{Caselli1997,gusdorf_sio_2008a,Jimenez-Serra08, Guillet2011}. Among these species, SiO stands out as a prominent tracer of shocks, with abundances enhanced by up to six orders of magnitude \citep[e.g.][]{Bachiller1991,Bachiller1998, Martin-Pintado92, Codella1999, Nisini2007}. 
\begin{figure}
\begin{center}
\includegraphics[width=0.9\textwidth]{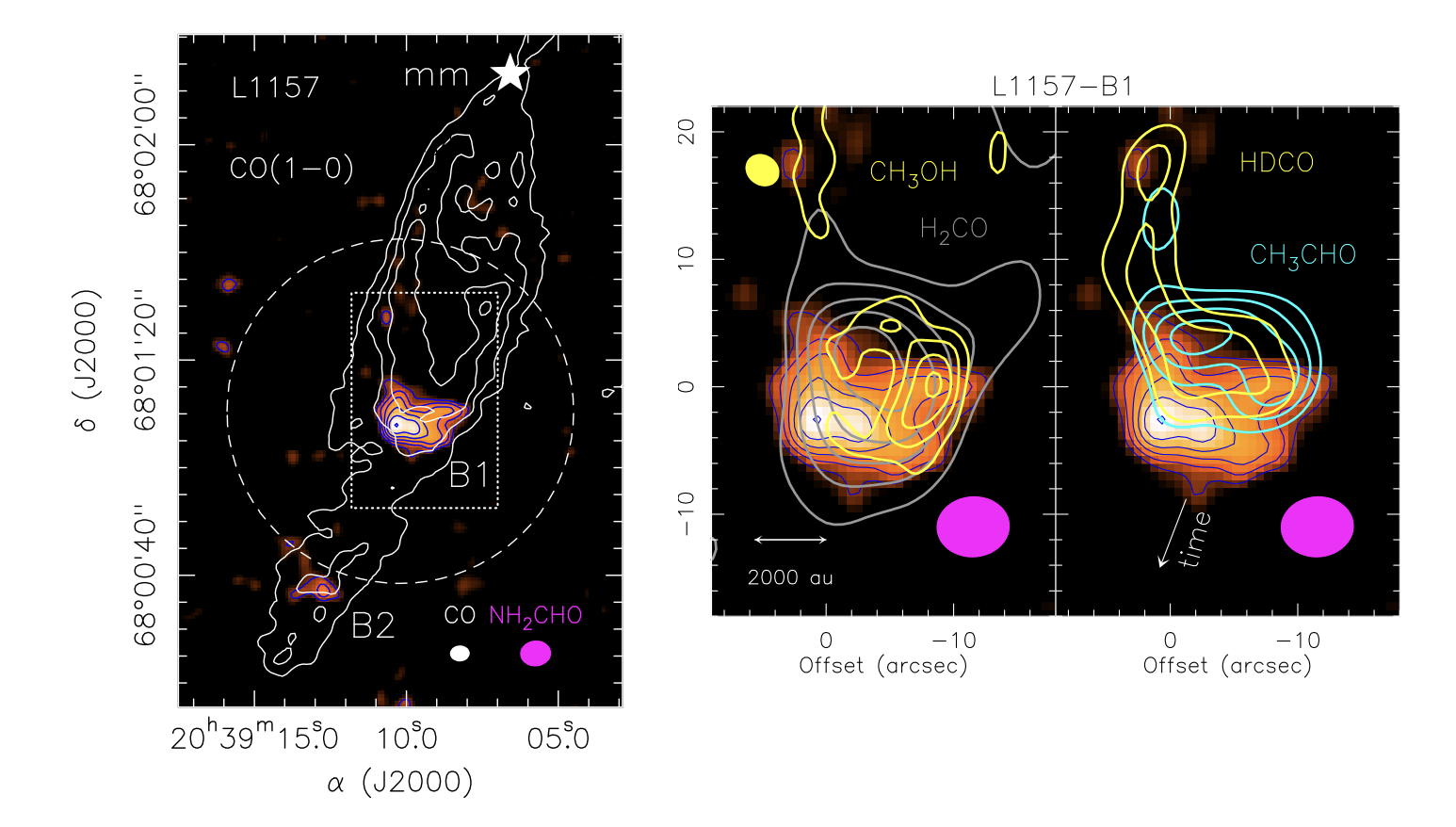}
\vspace{-15pt}
\caption{Chemical stratification in the L1157-B1 shocked region as traced by the IRAM-NOEMA. (Left) Maps of NH$_2$CO~(4$_{\rm 1,4}$–-3$_{\rm 1,3}$) and CO(1--0) emission (white contours; \citealt{Gueth1996, Codella2017}). (Right) Zoom-in of the B1 shocked region as traced by Formamide (colour), Methanol, Formaldehyde, deuterated Formaldehyde, and Acetaldehyde \citep{Benedettini2013,Fontani2014,Codella2015,Codella2017,Codella2020}. For each line image, synthesised beams are shown. Adapted from \citet{Ceccarelli2023}.} 
  \label{fig:l1157}
  \end{center}
\end{figure}
Similarly, Methanol (CH$_3$OH) was among the first iCOMs\footnote{interstellar Complex Organic Molecules: C-bearing species with $\geq$6 atoms and containing heteroatoms as Oxygen and/or Nitrogen \citep{Herbst2009,Ceccarelli2023}.} found to be enhanced (up to a factor $\sim$100) in shocks, tracing high-density and high-temperature regions (\citealt{Bachiller1997,Bachiller1998methanol,Bachiller2001,Jorgensen2004,Codella2025}). The intense radiation produced in shocks can substantially modify the chemistry of dense condensations that have not yet experienced any dynamic disturbances \citep[e.g.,][]{Girart2002, Girart2005, Viti2006, Tappe08}. Subsequent improvements in sensitivity and resolution enabled the detection of more complex species, such as Methyl formate (CH$_3$OCHO), Acetonitrile (CH$_3$CN), Formamide (NH$_2$CHO), and Ethanol (C$_2$H$_5$OH), especially toward the prototypical chemically-rich outflow driven by the L1157-mm protostar \citep[e.g.][]{Arce08,Codella2009,Sugimura2011,Yamaguchi2012,Lefloch2017,Lefloch2018, Lopez-Sepulcre2024}, as well as in other outflows associated with low-, intermediate-, and high-mass protostars \citep[e.g.][]{Palau2017,Holdship2019, Desimone2020solis}.\\
Single-dish telescopes (e.g., APEX\footnote{\url{https://www.eso.org/public/teles-instr/apex/}}, IRAM\footnote{\url{https://iram-institute.org/}} 30m) have been essential in revealing chemically enriched outflows. However, sensitive high-angular-resolution interferometric observations (with e.g., ALMA and IRAM NOEMA) are required to exploit the diagnostic power of molecular emission. Figure \ref{fig:l1157} shows the chemical richness of the L1157-B1 shocked clump. These observations enable resolving spatial structures and detecting chemical segregation within shocked regions -- critical for constraining the physical properties of jets and the formation pathways of iCOMs.\\
Works performed in the framework of the large programme IRAM NOEMA SOLIS\footnote{\url{https://solis.osug.fr/-THE-PROJECT-}} \citep{Ceccarelli2017} revealed a spatial chemical segregation in L1157-B1 and B2 (Figure~\ref{fig:l1157}) among several iCOMs \citep{Codella2017,Codella2020,Lopez-Sepulcre2024}. The prior knowledge of the outflow properties combined with the comparison between observations and astrochemical models allows the characterisation of the dominant formation routes of the detected iCOMs. Indeed, the observations provide the chemical stratification in a time-dependent structure, while the model provides the evolution of the molecular abundances in the post-shocked gas \citep{Giani2023}. This strategy has been extended to other sources (NGC1333-IRAS 4A, and OMC-2/3 FIR6 c-a) where high-resolution imaging has uncovered similar stratification \citep{Desimone2017,Bouvier2025}. Finally, ALMA observations of masers of Hydrogen recombination lines at sub-mm wavelengths have been instrumental in unveiling the kinematics and the structure of ionised jets and winds in massive objects (see Sect.~\ref{sec:SKA_SC3-4_atomicJet}). A striking example is the massive star MWC 349A, where the improved sensitivity and angular resolution with respect to previous telescopes allowed scientists to spatially resolve the inner $\sim$100 au of the ionised wind and jet, probing their velocity structure close to the launching zone \citep{Prasad2023, MartinezHenares2023, MartinezHenares2024}. Furthermore, the detection of a thermal hydrogen radio recombination line towards a protostellar jet \citep{JimenezSerra2011, SanchezMonge2025} opens the door for future discoveries that will enable the characterisation of the thermal jet's physical properties (e.g., density, temperature, ionisation fraction (\ionfrac), line-of-sight -- LOS -- velocities; see Sect.~\ref{subsec:RRLs}).\\
In summary, sub-mm observations, particularly when carried out at high angular resolution, are essential for dissecting the physical and chemical structure, the origin, and the evolution of molecular outflows and jets. While single-dish telescopes have been critical for discovering chemically active outflows, interferometers allowed us to resolve the spatial and velocity structure, disentangle chemical layers, and match observations with detailed astrochemical models. These efforts are crucial for using molecular emission as a diagnostic tool not only for the jet dynamics and shock physics, but also of the chemical evolution of protostellar environments. However, sub-mm observations have highlighted an intrinsic limitation that remains insurmountable at these wavelengths: the absence of crucial information regarding heavier carbon-bearing species, which are key for our understanding of how chemical complexity grows in the ISM (see also the chapter by \citealt{Bianchi01.2026.SKA} in this volume). Within the broader scope of scientific investigations supported by the SKAO, it is noteworthy that a number of relatively simple carbon chain molecules and cyanopolyynes, such as ethynyl (C$_2$H), cyclopropenylidene (c-C$_3$H$_2$), and cyanoacetylene (HC$_3$N), have already been successfully observed in the millimeter (mm) spectral range in the prototypical shocked region L1157-B1. The accumulated observational evidence \citep[e.g.][]{Benedettini07, Sugimura2011} has laid important groundwork for extending the chemical inventory toward more complex and massive carbon-bearing species.
These heavier molecules are expected to emit primarily at frequencies below 15 GHz, which aligns well with the capabilities of upcoming SKAO surveys. The presence of such species in the post-shock environments of astrophysical sources can offer unique diagnostic tools to trace the progressive chemical enrichment of the ISM. In addition, shocks can be used as laboratories to understand the chemical composition of the refractory part of dust cores, which remains largely unexplored, even after the launch of the JWST due to the limitations of mid-IR observations to the unambiguous identification of the IR bands of refractory material (e.g. \citealt{Kimura05}). Complementary to far-IR space missions \citep{Jimenez-Serra25}, the SKA telescopes have the potential to play a crucial role in the characterisation of the refractory part of dust grains - which is the material incorporated later on into cometesimals and planetesimals - through observations of the rotational lines of metal-bearing molecular species at cm-wavelengths (see Sect.~\ref{subsec:astromineralogy}).

\subsubsection{Centimetre-Decimetre telescopes}\label{sec:cm-telescopes}
Due to the severe angular resolution limitation of single-dish telescopes operating at cm wavelengths, almost all studies on YSO outflows in this range have been carried out with aperture-synthesis telescopes, notably mostly with the {\it Very Large Array} (VLA). However, several other facilities have contributed significantly to the field: the {\it Australian Telescope Compact Array} (ATCA) has been key in surveying southern regions inaccessible to the VLA; {\it eMerlin} provides the highest high angular resolution (50 mas at 5 GHz) for non-VLBI telescopes; The {\it Giant Metrewave Radio Telescope} (GMRT) has enabled high-sensitivity exploration at low frequencies ($\nu<1$~GHz), where non-thermal emission appears to be important. With the development of the VLA in the early 1980s, the high-resolution and sensitive observations revealed several low-mass YSOs exhibiting radio emission. Although these emissions were initially thought to be H{\footnotesize II} regions \citep{Haschick80}, it was quickly recognised that the radio brightness exceeded what would be expected based on the YSOs luminosity. Consequently, the emissions were associated with the mass loss of the YSOs \citep[e.g.,][]{Cohen82, Pravdo85, Snell85}. After these early studies, there have been numerous radio observations at GHz frequencies \citep[e.g.,][]{Andre87, Anglada92, Girart02, Reipurth04, Dzib2013}. 
The emission is usually weak, with flux densities in the [$\mu$Jy, mJy] range. These observations have revealed peculiar flat or mildly positive spectral indices ($S_\nu \propto \nu^{\alpha}$, with typically $\alpha\simeq 0.5$), and when resolved, they show extended and elongated structures in the same direction as the larger-scale optical jets and molecular outflows \citep{Curiel93, Anglada98}. It is now well established that cm free-free emission is associated with jets driven by YSOs of all stellar masses \citep{Anglada2018}.
All these characteristics can be understood with the free-free jet models by \cite{Reynolds1986}, and therefore they are usually referred to as "thermal radio jets" \citep[e.g. Cep~A HW2, ][]{Rodriguez1994}. The radio luminosity ($L_{\rm R}$) of  YSOs, is correlated with the strength of the outflow and with the YSO bolometric luminosity \citep[\Lbol;][]{Anglada2018}. This is a clear indication that the radio emission is associated with the outflow phenomenon. However, the most evolved YSO (mainly Class III) show non-thermal radio emission associated with the active stellar magnetosphere \citep[e.g.,][]{Andre1992, Gudel2002}. Both contributions (free-free emission from outflow and the non-thermal emission) can be found also in intermediate stages, such as late Class I or Class II \citep[e.g.,][]{Girart2004, Coutens19, Kaur2024}.
In addition to the radio emission around the youngest YSOs, a small number of HH objects also exhibit radio emission \citep[e.g.,][]{Anglada92, Marti1993, Masque2012}. The spectral index can be compatible with optically thin free-free emission \citep[e.g., HH 1--2,][]{Rodriguez2000}, or be slightly negative, indicating a non-thermal contribution \citep[e.g., HH~80-81-80N,][and Sect.~\ref{sec:SKA_SC1_Non-therm-Jet}]{Marti1993, Masque2012}.\\
Over the past thirty years, spots with negative spectral indices (signatures of non-thermal emission) have been detected along the route of YSO jets. These spots are typically associated with fast-moving radio knots that trace strong shocks where jets interact with dense ambient material \citep[e.g.,][]{Marti1995, Rodriguez2000, Curiel2006, RodriguezK2016, RodriguezK2022, Sanna2019}. 
Some of the clearest examples include the triple source in Serpens \citep[e.g.,][]{Rodriguez1989, RodriguezK2016}, HH 80–81 \citep[e.g.,][]{Marti1993, CarrascoG2010, RodriguezK2017, Vig2018}, Cep A \citep[e.g.,][]{Garay1996}, W3(OH) \citep[e.g.,][]{Wilner1999}, IRAS 16547–4247 \citep[e.g.,][]{Rodriguez2005}, HOPS 370 \citep[e.g.,][]{Osorio2017}, G035.02+0.35 \citep[e.g.,][]{Sanna2019} and DG Tau A \citep[e.g.,][]{Feeney-Johansson19}. Over the past decade, studies of large samples of YSOs revealed an increasing number of radio jet knots with negative spectral indices \citep[e.g.,][]{Purser2016, Rosero2016, Sanna2018, Kavak2021, Obonyo2025}, broadening the sample of objects to be considered potential sources of non-thermal emission.
Outflows have also been studied in maser emission using VLBI techniques, especially water masers \citep[e.g.,][]{Torrelles2011,  Moscadelli2005, Moscadelli2019,Vlemmings2006, Alves2012, Sanna2012, Carrasco-Gonzalez2015, Goddi2017}. These observations trace shocked outflowing gas very close to its launching zone and provide accurate measurements of the collimation and kinematic properties of the flow, in particular the signatures of rotation around the symmetry axis. These results have validated the widely adopted model of magneto-centrifugal acceleration for the launch of outflows \citep{Surcis2014, Moscadelli2022}.

\section{Non-thermal emission in jets} \label{sec:SKA_SC1_Non-therm-Jet}
\begin{figure}
\begin{center}
\includegraphics[width=0.95\textwidth]{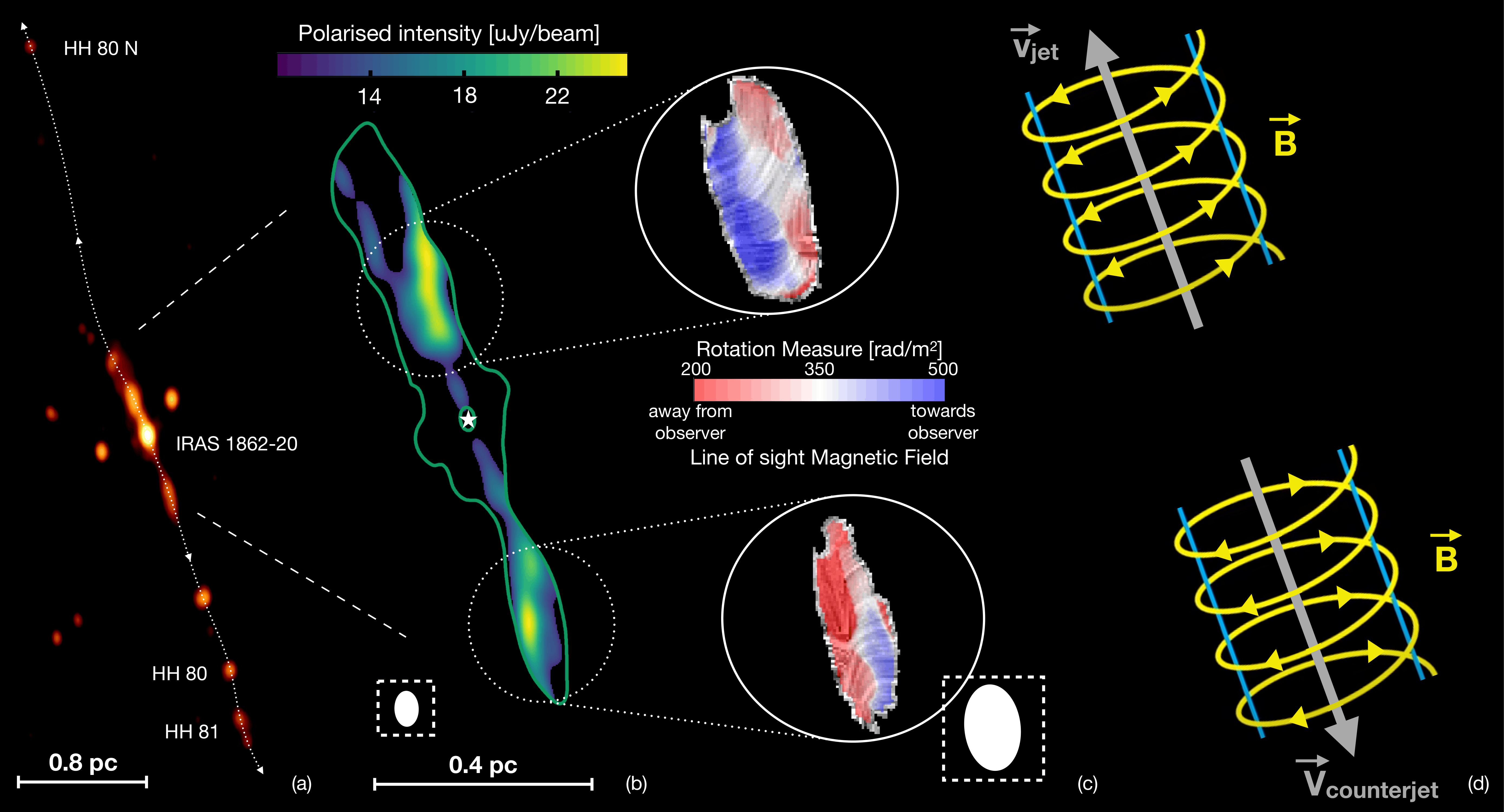}
  \caption{Full 5 cm view of the HH 80-81 jet. (a) Total intensity image of the radio jet, showing precession with dotted arrows; Herbig–Haro objects are labelled. The dashed lines indicate the region shown in (b). (b) Polarised emission is shown where the S/N is above 4 and total intensity S/N above 10. Green contours show total intensity at [10, 1000]$\sigma$ (with $\sigma$~=~3 $\mu$Jy beam$^{-1}$). Circles mark the area shown in panel c. Synthesised beam (10$^{\prime\prime}$ $\times$ 6$^{\prime\prime}$, PA = 4$^\circ$) is at lower left; 0.4 pc corresponds to 60$^\prime$. (c) Zoom into the lobes where the rotation measure (RM) was analysed (beam: 12$^{\prime\prime}$ $\times$ 7.5$^{\prime\prime}$, PA = 7$^\circ$, lower right). Overlaid are magnetic field streamlines on the RM map, with colour scale indicating LOS-magnetic field direction (red: pointing away, blue:  pointing toward observer). B-field and RM are shown for pixels with S/N $>$ 7 in Stokes I and polarised intensity. (d) Schematic illustrating the inferred 3D helical magnetic field configuration from polarisation analysis. Figure adapted from \cite{RodriguezK2025}.}
  \label{fig:RM-hh8081}
  \end{center}
\end{figure}

Non-thermal emission in protostellar jets is attributed to a minority population of relativistic particles accelerated in strong shocks. The observational evidence of non-thermal emission has also been addressed from a theoretical point of view. It has been shown that the interstellar flux of relativistic CR electrons is not sufficient to explain the synchrotron flux density observed in protostellar jet knots as well as near the protostellar surface.
Among the different mechanisms that can accelerate thermal particles to relativistic energies, such as stochastic Fermi acceleration, magnetic reconnection, and first-order Fermi acceleration \citep{Marcowith16}, the latter, also known as diffusive shock acceleration (DSA; e.g. \citealt{Bell78}), has been identified as the main mechanism for efficient acceleration at shock fronts present in both intermediate- and high-mass YSOs (\citealt{Padovani15,Padovani20}, see also \citealt{RodriguezK2016, RodriguezK2017} for an observational perspective).
The advantage of this theoretical approach is that it is possible to reproduce the observed flux densities by quantitatively constraining physical parameters such as volume density, ionisation fraction, temperature, flow velocity in the shock reference frame, and magnetic field intensity \citep{Padovani16}. In addition to non-thermal emission, the same model can be used to explain the origin of the enhanced ionisation rate (higher than the estimated range of $\sim$10$^{-18}$-10$^{-16}$~s$^{-1}$ due to Galactic cosmic-ray flux; e.g. \citealt{Sabatini20, Sabatini23, Socci24, Redaelli25}) necessary to reproduce the abundance of molecular species such as HCO$^+$ and N$_2$H$^+$ in L1157-B1 \citep{Padovani16} and cyanopolyynes in OMC-2 FIR 4 \citep[][see also Sect.~\ref{sec:SKA_SC5_molecules}]{Fontani17, Lattanzi23}. A double signature of the presence of a local source of energetic particles (non-thermal emission and high ionisation rate, \citealt{Padovani21}) has therefore been proposed, and SKAO is the key instrument for this type of analysis.
So far, in one object only (HH 80-81; Fig.~\ref{fig:RM-hh8081}), non-thermal knots were confirmed to be due to synchrotron emission by direct detection of linearly polarised emission \citep{CarrascoG2010}. More recently, \cite{RodriguezK2025} performed the first rotation measure analysis of polarised synchrotron emission in the same jet, revealing a 3D helical magnetic field (B-field) structure. This breakthrough paves the way for further studies of the magnetic field strength and morphology in protostellar jets in the near future.\\
However, whether non-thermal emission is common in protostellar jets remains a key question, directly linked to the identification of the underlying mechanisms that could enable efficient particle acceleration across different protostellar masses. The SKAO will provide a transformative improvement to this research line and greatly increase the number of studies of polarised emission and related B-field in a large sample of protostellar jets exhibiting signs of non-thermal emission. In general, negative spectral indices are found toward jet lobes at relatively large distances ($>$3000~au) from the protostar, likely associated with shocks interacting with the ambient medium. As such, radio sources showing extended lobes are good candidates for SKAO investigations aimed at the study of particle acceleration and synchrotron emission. However, there are cases where non-thermal emission has been proposed to arise at small (projected) distances ($\sim$100 au) from the YSO \citep[e.g., NGC 1333 IRAS 2B;][]{Tychoniec2018}.

\subsection{Synchrotron emission and linear polarisation}
Synchrotron emission is a powerful diagnostic tool for probing B-fields in astrophysical jets, as it reveals both their presence and their properties. As linearly-polarised radio waves propagate through magnetised plasma, their polarisation plane rotates due to Faraday rotation. This rotation is proportional to the Faraday depth, which encodes information about the B-field component along the line of sight. Fitting a Faraday rotation model to the observed Stokes parameters, the intrinsic (emitted) polarisation can be recovered, thereby enabling the 3D B-field structure to be inferred. In an ideal uniform magneto-ionic external medium - a Faraday screen - the Faraday depth equals the rotation measure (RM), which characterises the simplest case where the polarisation angle varies linearly with wavelength squared: $\chi(\lambda^2) = \chi_0 + \mathrm{RM} \lambda^2$, where $\chi_0$ is the source’s intrinsic polarisation angle. This indicates pure rotation without depolarisation or changes in the intensity of the polarisation vector. This analysis enables RM and $\chi_{0}$ maps to be created. Rotating polarisation vectors by 90$^{\circ}$ yields the B-field orientation in the sky plane, while RM values provide information on the B-field component in the LOS direction. These techniques are well established for relativistic jets from Active Galactic Nuclei \cite[e.g.,][]{Pasetto2021} and have been applied to the protostellar jet HH\,80-81 \citep{RodriguezK2025}, powered by the massive protostar IRAS\,18162-2048 (Fig.~\ref{fig:RM-hh8081}a). The advantage in studying young stellar object jets is that both the jet and counterjet lobes have comparable signal-to-noise, providing a 3D description of the B-field at opposite positions with respect to the star, and allowing us to relate it with the  disc kinematics itself. Notably, HH 80-81 remains the only YSO jet with detected linearly polarised synchrotron emission so far.\\
In HH~80–81, the 3D B-field structure within the jet was reconstructed through a radio polarimetric study (see Fig.~\ref{fig:RM-hh8081}b) based on deep VLA observations \citep{RodriguezK2025}. The method relies on analysing the polarisation angle and degree as a function of wavelength across the observed radio band, which requires highly sensitive data to detect polarised emission in narrow sub-bands. Assuming a simple Faraday screen model, the wavelength dependence of Stokes Q and U is simultaneously fitted, allowing recovery of the intrinsic polarised emission properties and reconstruction of the B-field geometry. Figure~\ref{fig:RM-hh8081}c shows the plane-of-sky magnetic field ($B_\perp$) streamlines overlaid on the RM map, where red and blue indicate LOS field components ($B_\parallel$) directed away from and toward the observer, respectively. From these results, the authors inferred a helical configuration of the B-field, with $B_\parallel\sim$~0.1~mG consistent with the equipartition-derived plane-of-sky value of $B_\perp\sim$~0.2~mG  \citep{CarrascoG2010}, giving an average strength of $B=\sqrt{B_\perp^2+B_\parallel^2}\sim$0.22~mG. A schematic of the inferred helical field is shown in Fig.~\ref{fig:RM-hh8081}d. The field winding follows the disc rotation (counterclockwise as seen from Earth; e.g., \citealt{CarrascoG2012}; \citealt{FernandezL2023}), but its orientation appears to be independent of the rotational motion (i.e., RM gradients in the jet and counterjet exhibit clear opposition). We stress this significant result because the orientation of the magnetic fields may depend on the mechanism generating them.

\subsection{Observational prospects with the SKA-Mid}
The SKA-Mid\footnote{\url{https://www.skao.int/en/explore/telescopes/ska-mid}\label{webSKAmid}} unprecedented sensitivity will expand the sample of YSOs exhibiting non-thermal radio jets and will make it possible to extend linear polarisation studies to protostellar jets beyond the single case identified so far (i.e. HH80-81). This will enable the investigation of B-field properties in a broader population of protostellar jets. Here, we provide an estimate of the time needed to detect non-thermal emission in YSO jets with the SKA telescopes, based on the parameters reported in the literature. The expected polarisation fraction in synchrotron-dominated regions ranges from 5\% to 30\%, depending on the B-field alignment and observational wavelength. According to the literature, typical flux densities of radio jets and knots in HH objects at cm-wavelengths ($\sim$10 GHz) are in the range $\sim$0.1 to a few mJy (e.g, \citealt{Anglada2018}). At lower frequencies, the emission is expected to be higher for non-thermal sources: e.g., a source with a flux density of 0.2 mJy at 10 GHz is expected to have a flux density of $\sim$1 mJy at the frequencies covered by SKA-Mid Bands 1 (0.80 GHz) and 2 (1.31 GHz), assuming a spectral index of $-$0.6 \citep[e.g.,][]{RodriguezK2016, Osorio2017, Vig2018}. Assuming a conservative polarisation fraction of 5-10\%, we estimate the expected polarised flux density of a non-thermal radio jet to be on the order of 50-100 $\mu$Jy.\\
Using the SKAO Sensitivity Calculator\footnote{All integration times provided in this Chapter are obtained using the SKAO Sensitivity Calculator (\url{https://sensitivity-calculator.skao.int/}) updated as of October 2025. The inclusion of MeerKAT antennas for Band~5 calculations follows the methodology discussed by \cite{Traficante01.2026.SKA} in this volume.} in continuum mode (AA4 configuration), we estimated the integration times needed to reach an rms noise level of 2 $\mu$Jy/beam over the full bandwidth in Band~1 and 2 at a declination of $-5^{\circ}$ (Orion). To achieve a synthesised beam of $\sim 1''\times 1''$, we assumed Briggs weighting with robust = 0 in Band~1 and robust = 1 in Band~2. The requested rms is reached in $\sim$1.4~h in Band 1 (435~MHz bandwidth), and $\sim$11~min in Band 2 (720~MHz bandwidth). Note that these estimates are intended to detect the presence of a non-thermal emission component with the purpose of identifying new synchrotron emitter candidates.\\
In-depth studies will be crucial to investigate the 3D-structure of the B-field, requiring the source to be angularly resolved, and in the following, we estimate the sensitivity and integration times needed with SKA-Mid. 
As a suitable example to illustrate this calculation, we consider the case where the angular size of the emission (which can vary with wavelength) is sampled by 5 beams, corresponding to the source solid angle. Following \cite{RodriguezK2025}, we must study the in-band, frequency dependent, polarisation behaviour. To do this, we need observations sensitive enough (e.g., S/N$>$7) to detect a clear signal of linearly polarised emission in a number of sub-bands within a given band. For a total polarised flux density of 50-100 $\mu$Jy, we would expect an average intensity of 10-20 $\mu$Jy~beam$^{-1}$. We estimate the integration time to detect polarised emission at a S/N $\sim$ 7 (i.e. rms~$\sim$~1.5-3~$\mu$Jy~beam$^{-1}$) per sub-band for sources with peak intensities of 10-20 $\mu$Jy~beam$^{-1}$, dividing the full bandwidth into 20 sub-bands of 36 MHz (Band 2) and $\sim$21 MHz (Band 1). With a synthesised beam of $\sim 1''$ (refer to the parameters above), reaching an rms of 3~$\mu$Jy beam$^{-1}$ requires $\sim 1.7$ h in Band 2 and $\sim 12.5$ h in Band 1. To achieve a deeper sensitivity of 1.5~$\mu$Jy beam$^{-1}$, the required integration times increase to 6.7~h and 50~h, respectively.\\
This investigation would be feasible even during the SKA telescopes science verification phase, as suggested by analogous calculations performed for the AA* configuration. With a $\sim 3''\times 3''$ synthesised beam and considering 10 sub-bands, polarised emission in Band 2 can be detected at S/N $\sim$7 per sub-band for resolved sources (solid angle $\sim$5 beams) with intensities of 10-20 $\mu$Jy~beam$^{-1}$ requiring 2 to 8~h of integration. In Band 1, detecting a 20 $\mu$Jy~beam$^{-1}$ source at S/N $\sim$7 per sub-band requires about 14~h, while for a fainter source of 10 $\mu$Jy~beam$^{-1}$, the required time increases by a factor of four.

\section{Thermal emission in protostellar jets}\label{sec:SKA_SC2_ionizedJet}
Over the past decades, significant progress in characterising the thermal emission of radio jets from YSOs has been driven by radio interferometric facilities working in the centimeter range. These instruments enabled sub-arcsecond imaging of jet morphologies, detection of variability, and precise measurements of spectral indices \citep[see Sect.~\ref{sec:cm-telescopes} and ][for a review]{Anglada2018}. In parallel, deep surveys have established a robust empirical correlation between the $L_{\rm R}$ of jets and \Lbol~ of their driving sources that holds for stars with high luminosities down to very-low luminosity objects (VeLLOs), and even proto-brown dwarfs (see Sect.~\ref{subsec:LrLbol}). This relation underscores the universality of mass-ejection processes across the protostellar population.\\
Despite the tremendous progress made in last years, the current facilities impose important limitations to fully understand thermal radio jets. Their angular resolution and sensitivity constrain our ability to resolve the innermost jet-launching regions, specially at low luminosities or in distant and crowded star-forming regions. Moreover, frequency coverage is often insufficient to disentangle free–free emission from possible non-thermal contributions or to trace the full spectral turnover from optically thick to thin regimes. These challenges highlight the need for next-generation radio arrays -- such as the SKA telescopes -- to fully characterise the physics of thermal radio jets.

\subsection{Launching and collimation mechanisms}\label{subsec:lanchingjet}

It is not yet clear whether the launching and collimation mechanisms of protostellar jets are universal across all stellar masses and evolutionary stages, or if they vary as a function of protostellar \massacc, mass, or environment. In solar-type objects, observations suggest that collimation takes place very close to the protostar, often within $\sim1$~au. An example is HL\,Tau, a Class~II YSO, where cm-continuum emission, with a size of only 3~au, is elongated along the same direction than the optical jet \citep{Carrasco-Gonzalez2019}. In the intermediate-mass regime, high angular resolution observations reveal the coexistence of two ionised components: one poorly collimated wind and a second, well-collimated jet originating in the immediate vicinity of the protostar \citep{RodriguezK2022,Carrasco-Gonzalez2021}. Only magnetic fields are capable of providing collimation on such short distances, thus this observational evidence is a strong indication of an intrinsic magnetic drive for the launch of jets, as, for example in the models of magnetised disk winds (see \citealt{Ray2021NewAR..9301615R} for a review). However, for some massive objects, there is evidence of collimation occurring at larger distances, beyond 10~au from the central source \citep{Carrasco-Gonzalez2015, Carrasco-Gonzalez2021}. In massive star-forming environments, external collimation mechanisms could play a significant role. Laboratory experiments and theoretical work \citep{Albertazzi2014} suggest that poorly collimated winds could be gradually focused as they interact with a dense surrounding medium and/or a strong ambient magnetic field (likely to be present in high-mass star-forming regions). Understanding whether collimation is an intrinsic property of the launching mechanisms or the result of external confinement is therefore a crucial open question in the study of protostellar jets and outflows. 

\begin{SCfigure}
    \centering
\includegraphics[width=0.56\textwidth]{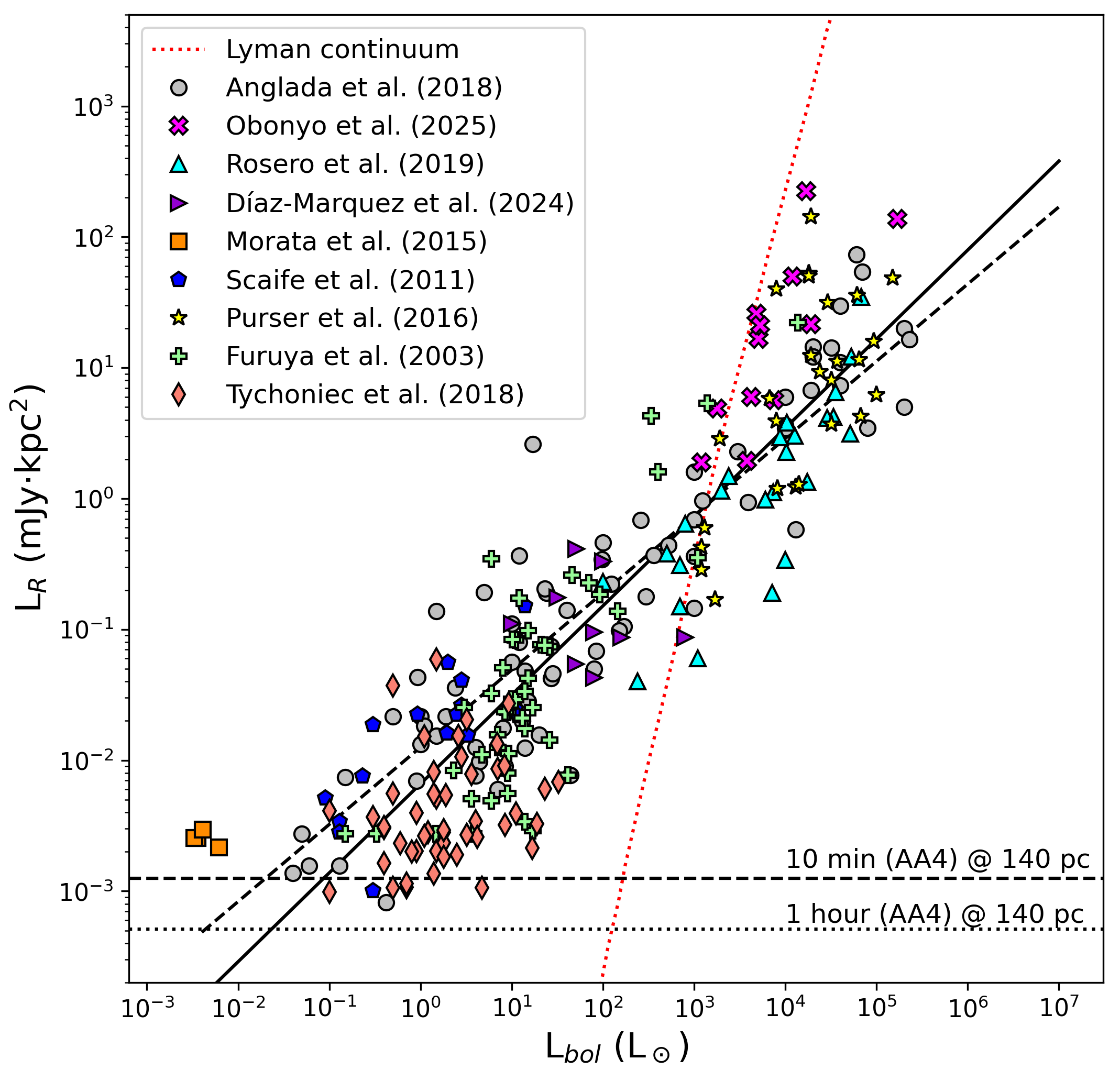}
\caption{$L_{\rm R}$-\Lbol~ correlation using data collected from the literature. The dashed black line corresponds to the fit done to all radio jets in \cite{Anglada2018}, while the solid black line is the fits using the whole sample. The red dotted line depicts the expected $L_{\rm R}$ associated with the Lyman continuum flux of \hii~regions powered by stars of different \Lbol. The black dashed horizontal line marks the 5$\sigma$ rms of SKA-Mid Band~5a in AA4 (without MeerKAT antennas) after 10 minutes of integration time ($\sim0.0012$~mJy\,kpc$^2$) for sources located at 140~pc. The black dotted horizontal line represents the one after 1~hour ($\sim0.00051$~mJy\,kpc$^2$).}\label{fig:lrad-lbol} \nocite{Anglada2018, Furuya2003, AMI2011, Tychoniec2018, Morata2015, Diaz-marquez2024, Rosero2019, Purser2016, Obonyo2025}    \label{fig:lrad-lbol}
\end{SCfigure}

\subsection{The radio-bolometric luminosity relation: ejection efficiency}\label{subsec:LrLbol}

A key finding in the study of protostellar jets is the established correlation between the $L_{\rm R}$ of the jets and \Lbol~ of their driving protostars (see Figure~\ref{fig:lrad-lbol}). This correlation provides a valuable framework for estimating the expected flux across a wide range of masses. However, observational challenges exist at both ends of this correlation. For massive YSOs, the radio emission associated with the jet is several orders of magnitude weaker than the one expected from photoionization. In the crowded environments of the high-mass star-forming regions intense radio continuum emission associated with \hii~regions complicates the detection of the jets. Thus observations with very high dynamic ranges are essential. On the other hand, jets from very low-mass objects are intrinsically very weak. Detecting these faint sources requires higher sensitivities. Furthermore, observations across all mass ranges are strongly affected by distance. Currently, the study of radio jets, particularly for low-mass YSOs, is limited to the nearest star-forming regions ($<$1 kpc).\\
VeLLOs, such as brown dwarfs, exhibit an unusual excess in $L_{\rm R}$ relative to the established correlation for massive and low-mass YSOs \citep{Ellerbroek2013,Palau2024}. This behaviour could be interpreted as these very low-mass objects transforming accretion into ejection more efficiently, thus showing a high value of ratio between the mass-loss (\massloss) and mass-accretion (\massacc) rates. This interpretation is supported by observations of solar-type Class~II sources, where an increase in efficiency has also been observed, with lower accretors exhibiting a higher ratio between ejection and accretion \citep{Rota2025}. However, the derived \massloss~strongly depends on the assumed \ionfrac~of the jet. A value of $\chi_{\mathrm{e}}\approx10$~\% is typically adopted for all jets regardless of the mass of the driving source. We note, however, that such a value for $\chi_{\mathrm{e}}$ derives from the few cases in which this difficult measurement has been possible, that is for a handful of solar-mass T Tauri stars \citep[e.g.,][]{Bacciotti1999,Hartigan2007}. Recent measurements in massive jets yielded a value of 5\%-12\%, \citep{Fedriani2019}.

\subsection{The role of SKA-Mid: Probing the Jet launching mechanism and proper motion}\label{subsec:SKAmid_launching}
The advent of the SKAO will open a transformative window for the study of thermal radio jets from protostars and YSOs. In particular, the superb angular resolution ($\sim4$~au for a distance of 200~pc) and sensitivity ($\sim4$~$\mu$Jy in 1 hour) of SKA-Mid in Band~5b will address long-standing questions on the launching and collimation mechanisms of jets.\\
A key motivation for a better understanding of the  connection between \massloss~and \massacc~is to refine and extend the empirical $L_{\rm R}$-\Lbol~ relation discussed in Sect.~\ref{subsec:LrLbol} (see Fig.~\ref{fig:lrad-lbol}). This relation currently spans several orders of magnitude, but it is poorly constrained at its extremes. SKA-Mid will enable detections of statistically significant samples towards the low-luminosity end (i.e., VeLLOs and proto-brown dwarfs) in regions of our solar neighbourhood. As shown in Fig.~\ref{fig:lrad-lbol}, already with 10 minutes of integration time in Band~5a the 5$\sigma$ sensitivity limit is well bellow the current measurements with existing facilities, and after 1 hour of integration time, the detection limit decreases by almost one order of magnitude. Recently, \citet{Perez-Garcia2025}, presented the SUbstellar CANdidates at the Earliest Stages (SUCANES) database, which contains 174  objects classified as potential very young substellar candidates. This sample offers an excellent example of the scientific exploitation of SKA-Mid Bands 5a and 5b. Moreover, the SKA-Mid long baselines for Band-5 gives an angular resolution $<0.1''$.  Upon the commencement of observations with SKA-Mid, a temporal baseline ranging from over 30 years will exist for high-angular resolution radio studies. This duration will facilitate the measurement of proper motions in the closest star-forming regions, such as Orion. On one hand, SKA-Mid will allow studying the kinematics of radio jets, overcoming resolution limits of previous measurements ($\sim$ 10 to 100~mas~yr$^{-1}$) of proper motions in radio knots from jets \citep{Rodriguez1989, Curiel93, Marti1998, Curiel2006, Masque2015, RodriguezK2022}. On the other hand, it will be possible to study the orbital properties of radio YSO binaries, and derive their masses \citep[e.g.,][]{Curiel2003, Loinard2003, Diaz2022}, as well as the kinematics of YSO clusters (e.g., \citealt{Dzib2017}; see Sect.~\ref{subsec:propermotion}). Note that studies of proper motions with current interferometers are not only very scarce, but they also can suffer from ``astrophysical noise''. In some cases, radio knots appear isolated and far from the central protostar \citep[e.g.,][]{RodriguezK2016, Carrasco-Gonzalez2021}, which makes it relatively easy to measure their proper motions. However, in some cases, especially when approaching the central protostar, radio knots appear embedded in extended emission which also changes with time \citep[e.g.,][]{Marti1995, Maureira20, Hernandez-Garnica2024}. This could lead to greater uncertainties in the jet's velocity or the source's proper motion. This effect which is important in current interferometers with a low number of antennas, could be highly mitigated thanks to the much higher image fidelity of the SKA-Mid.

\subsection{The role of SKA-Low: Optically thick emission}
The estimate of \massloss~ requires reliable determinations of the electron density (\nEL) and the ionised mass (\Mion) in the flow. These quantities can, in principle, be obtained from the observed radio flux density at a given frequency if we know the optical depth of the emission at that frequency. However, thermal radio jets at cm-wavelengths typically display spectral indices in the range 0.1--1 \citep[see e.g.,][]{Anglada2018}, indicative of  partially optically-thin emission. As a result the exact value of the optical depth remains uncertain. A common approach is to assume that the measured spectral index remains constant up to
the frequency at which the emission becomes totally optically thin, and where \massloss~could be safely determined~\citep[e.g.,][]{Anglada2018}. 
Unfortunately, the transition to total optical thinness happens at frequencies $>$20 GHz, where thermal dust emission from the disc and/or envelope often dominates the spectrum.\\
An alternative approach is to estimate \massloss~at the frequency at which the free-free emission becomes fully optically thick. This transition is expected to occur
at low frequencies, where contamination from dust is negligible. It is then possible to model the Spectral Energy Distribution (SED) and obtain \nEL, \Mion~and \massloss. However, the transition to optical thickness occurs around $\sim$100~MHz, where free-free emission is expected to be extremely weak. So far, this study could only be performed in the bright YSO T~Tau \citep[][]{Coughlan2017} through LOFAR observations. In this object, an evolved (Class II), low-mass object, the  transition to optically thick emission takes place at a very low frequency of $\sim$150 MHz, with a flux density of $\sim$2~mJy. However, T~Tau is a nearby ($\sim1$40~pc) and its jet is very bright, which makes it accessible to such studies.\\ 
SKA-Low\footnote{\url{https://www.skao.int/en/explore/telescopes/ska-low}\label{webSKAlow}} will open the possibility of extending this study to other objects. We can estimate the feasibility of detecting optically thick emission with SKA-Low at 200 MHz from Figure \ref{fig:lrad-lbol} (where $L_{\rm R}$ is usually measured at 5 GHz). We also assume a median spectral index of 0.45 \citep[][]{Anglada2018}. An integration time of just $\sim$1~hour with SKA-Low (AA4) yields an rms of $\sim$300$\mu$Jy~beam$^{-1}$,  corresponding to $L_{\rm R} = 6\times$10$^{-3}$~mJy~kpc$^2$ at 200 MHz ($\sim$0.03~mJy~kpc$^2$ at 5 GHz). This rms will allow us to detect optically thick free-free emission from jets driven by sources with \Lbol~$>10~L_\odot$. Detecting sources down to $1~L_\odot$ would require an integration time of $\sim$5~h, and hundreds of hours to cover the full \Lbol-range in Fig.~\ref{fig:lrad-lbol}. Note that the frequency at which emission becomes optically thick depends on density, meaning that denser jets are expected to be brighter at higher frequencies. Crucially, this part of the spectrum remains completely unexplored for protostellar jets.\\ 
The SKA-Low (AA4) angular resolution at 200 MHz will be $<3.5''$. While observations at 5~GHz typically resolve jets at subarcsecond scales, the expected size of free-free emission at 200~MHz is larger due to the frequency dependence of the optically thick region; for instance, in  T~Tau the jet emission was resolved at 150 MHz with a beam size of $\sim6''$.
The expected jet sizes ($\theta_{\rm jet,\nu}$) at 200 MHz can be estimated using the \cite{Reynolds1986} jet model, according to which $\theta_{\rm jet, \nu} \propto = \nu^{\alpha-1.3}$ \citep[see also][]{Anglada2018}. Assuming $\alpha$=0.45, a jet with $\theta_{\rm jet,5GHz} = 1''$ will have a size of $\theta_{\rm jet,200MHz} =15''$, and will be suitably resolved with the SKA-Low, up to a distance of 140 pc. 

\section{The atomic and ionised YSO jet components} \label{sec:SKA_SC3-4_atomicJet}
\subsection{Radio recombination lines (RRL) and \hi~emission}\label{subsec:RRLs} 
\begin{figure}
    \begin{center}
    \includegraphics[width=1\textwidth]{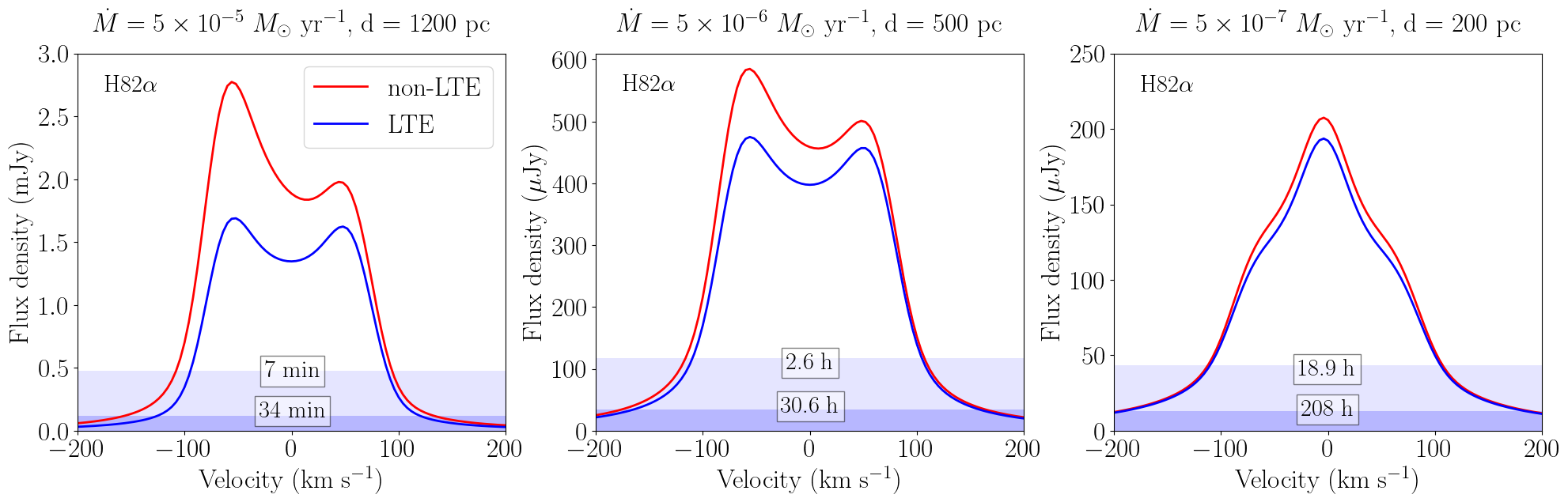}
    \caption{MORELI predictions of the H82$\alpha$ RRL at 11.71 GHz. We show predictions for different combinations of wind \massloss~and $d$ of the system, starting from the case of MWC\,349A (left panel). The pale-blue region shows the 5$\sigma$ detection with SKA-Mid (see text), while the dark blue corresponds to the 5$\sigma$ detection with SKA-Mid (AA4) of the line intensity at a velocity of 100 km~s$^{-1}$. Integration times from the SKAO Sensitivity Calculator are shown.}
    \label{fig:morelipredictions}
    \end{center}
\end{figure}

Hydrogen Radio Recombination Lines (RRLs) present an excellent opportunity to retrieve the kinematics of the ionised winds close to their launching sites, before they have interacted with the surrounding medium \citep[][see also the dedicated chapter by \citealt{Guidi01.2026.SKA} in this volume]{Anglada2018}. Interferometric observations of RRLs from massive systems trace jets \citep{JimenezSerra2011,Prasad2023}, MHD and photoevaporative (PE) winds \citep[e.g.][]{BaezRubio2014,Zhang2019}, and warped discs \citep{JimenezSerra2020}. Some of these sources present maser-amplified RRLs \citep[e.g. the massive star MWC\,349A,][]{MartinPintado1989}. Hence, non-LTE radiative transfer tools such as the MORELI code \citep{BaezRubio2013} are employed to characterise the launching sites and velocity structure of these winds  \citep{BaezRubio2014,MartinezHenares2023,MartinezHenares2024}. The detection of a thermal radio recombination line in the protostellar jets powered by Cepheus A HW2 (\citealt{JimenezSerra2011}) and W75N\,(B)-VLA3 \citep{SanchezMonge2025} offered a direct way to measure the jet's physical properties (e.g., \nEL, electron temperature, \ionfrac, and LOS-velocity). Despite these promising results, RRLs have not been detected yet in low-mass systems, with the exception of the externally-irradiated proplyds \citep{Boyden2025}. The radio continuum free-free emission detected in some low-mass stars suggests the presence of ionised winds from their discs, which would be confirmed by the presence of RRLs \citep[e.g.][]{Pascucci2012,Macias2016}. The main challenge comes from the fact that RRLs are expected to present low intensities under LTE conditions at centimeter wavelengths: i.e. only a few percent above the free-free radio continuum \citep[e.g.][]{Pascucci2018}. Although the RRL intensities increase as $\propto \nu^\delta$ with $\delta = {1.1}$, at mm-wavelengths the dust continuum may hinder their detection. The high sensitivity of SKA-Mid Band 5 observations provides, therefore, an ideal opportunity for the  detection of these lines across the full range of protostellar masses. Figure \ref{fig:morelipredictions} shows predictions of line intensities for a range of \massloss~and distances (d), and considering LTE and non-LTE effects. Using the SKAO Sensitivity Calculator for the AA4 configuration in Band~5b and a (velocity) spectral resolution $\sim$4 km~s$^{-1}$, we find that integration times for detection range from less than an hour to about twenty hours, depending on the source distance. This means that the detection of RRLs with the SKA-Mid will be feasible during the Science Veriﬁcation in AA$^*$ configuration, even toward distant low-mass YSOs. 
From these observations, the RRLs spatial distribution may be recovered by obtaining spectro-astrometric Gaussian centroids \citep[see e.g.][]{MartinPintado2011,Zhang2017,JimenezSerra2020, SanchezMonge2025}, which will allow the separation of the jet and wind contribution from the UC\hii~regions. The much higher velocities (of $\sim$100~km~s$^{-1}$) reached by jets and winds will also allow them to be distinguished from \hii~regions (whose expansion velocities are $<$20~km~s$^{-1}$). RRLs emitted at the frequencies covered by the SKA telescopes trace the outer, less dense regimes of the ionised disk and wind in comparison to the high-frequency lines covered by ALMA. This complementary view will allow us to retrieve the three-dimensional physical structure and kinematics of the wind and jet, and to test the MHD or PE origin of the outermost ejection. For the faintest lines, stacking of multiple RRLs will provide the required line intensities to prove the presence of ionised winds. The combination of these deep observations with non-LTE radiative transfer modelling will be crucial to investigate the origin of the jets and winds launched by YSOs.\\
At IR-wavelengths, high angular resolution observations with the JWST have shown that H recombination lines are not only associated with the accretion shock at the stellar surface but also exhibit considerable emission along the jet \citep[e.g.,][]{Harsono23,Federman24}, such as in the case of the very low-mass Class 0 protostar IRAS 16253–2429 \citep{Narang24}. The ionic jet component at these wavelengths is typically observed from the emission of forbidden lines such as [Fe II], [Ne II], and [Ar II], which provide a powerful probe of the hotter, more highly excited components of the outflow \citep{Tychoniec24,Garatti24,vanDishoeck25,Narang25}. However, in several YSOs, the large thickness of the parent envelope results in high dust extinction at the infrared. This, combined with the poorer spatial resolutions of the JWST with respect to interferometers, prevents the study of the jet at its launching position. In this context, the SKA telescopes will complement the JWST by enabling observations much closer to the jet base and providing velocity-resolved analysis, which will overcome the $\geq80$ km s$^{-1}$resolution limit of JWST.
Finally, while RRLs trace the ionised component of the jet and wind, the neutral atomic component often dominates over the molecular and ionised gas \citep[e.g.][]{Bally2016}. \hi~is the dominant component of the neutral jet material  \citep{Lizano88}, and accurately constraining the total mass, \massloss, and momentum carried by \hi~is essential for building a complete picture of the mass/energy budgets of these flows. However, \hi~emission at 21~cm from the neutral jet is often faint and can be contaminated by galactic \hi~emission along the same LOS. Despite these challenges, observatories such as {\it Arecibo} \citep{Lizano88}, the VLA \citep{Russell92, Giovanardi00} and now the {\it Five-hundred-meter Aperture Spherical radio Telescope} (FAST; \citealt{Li22}) show extended, high-velocity \hi~emission associated with outflows. These observations generally lack the spatial resolution needed to isolate the launching region, often resulting in confusion between overlapping flows from multiple protostellar sources and unrelated high-velocity clouds. The enhanced sensitivity and angular resolution of the SKA telescopes will enable the next major advancement in detecting and studying these neutral winds.

\subsection{Proper-motion of thermal/non-thermal knots}\label{subsec:propermotion}
Measuring jet velocities is essential to constrain theoretical models of jet launching and collimation (see Sect.~\ref{subsec:lanchingjet}). Velocities are combined with jet density to derive the \massloss~estimates and, in turn, the \massloss/\massacc~ ratio and the momentum flux (outflow force). All these parameters are key physical diagnostics for theoretical models \citep[e.g.][]{Frank2014, Ray2021NewAR..9301615R}. Achieving a complete dynamical picture of protostellar jets requires measuring their full 3D-velocity structure. While the precision of LOS velocity measurements of protostellar jets in the IR has improved significantly in recent years (e.g., $\sim$5–10 km~s$^{-1}$ with JWST/MIRI), it still remains moderate. The SKAO will enable the detection of RRLs in radio jets \citep{Anglada2015}, thereby providing even more precise measurements of the LOS velocity component. When combined with proper-motion measurements (i.e., the velocity components in the plane of the sky), this will allow us to reconstruct the full 3D velocity structure of the jet. The latter measurement relies on multi-epoch observations to find the proper motions of thermal and non-thermal continuum. These studies indicated  tangential velocities from $\sim$100 to $\sim$1000~km~s$^{-1}$ \citep[e.g.,][]{Marti1995, Marti1998, Rodriguez2000, Chandler2005ApJ...632..371C, Curiel2006, Masque2015, RodriguezK2016, RodriguezK2022, Osorio2017, Anglada2018}.\\
While current facilities can measure proper motions in only a few bright jets, the combination
of high angular resolution, wide bandwidth, and large collecting area offered by SKA-Mid will enable, for the first time, routine 3D-velocity measurements of faint (few mJy) knots (see Sect.~\ref{subsec:SKAmid_launching}). Starting from its first light in Science Verification, SKA-Mid will open up a new regime of kinematic studies of jets. For instance, in Band 5b ($\nu_0$~=~11.8~GHz; $\Delta\nu$~=~5 GHz) and 5a ($\nu_0$~=~6.55 GHz; $\Delta\nu$~=~3.0~GHz), thermal jets can be detected with an rms noise of about 1~$\mu$Jy$/$beam in $\sim$30 min for a source in Orion ($d$=390~pc; \citealt{Tobin2020}) using the AA4 configuration (with only 15-m antennas) and a synthesised beam of $\sim$0.1$''\times $0.1$''$ (Briggs; robust=0). Nonthermal knots from the same source can be detected in Band 2 ($\nu_0$~=~1310~MHz; $\Delta\nu$~=~720~MHz), with an rms noise of about 2~$\mu$Jy$/$beam in $\sim$30 min, using the AA4 configuration (including MeerKAT antennas) and a synthesised beam of $\sim$0.7$''\times$0.6$''$ (Briggs; robust=0). 
The 3D-velocity vector of each knot provides both the jet inclination and the true space velocity of the ejecta. This complete kinematic information enables: ($i$) the identification of acceleration and collimation through velocity gradients; ($ii$) refine estimates of \massloss~and momentum flux of the jet \citep[][]{Reynolds1986}; ($iii$) the detection of rotation or precession of jets via asymmetries in the velocity field; and ($iv$) to distinguish between different launch models by getting closer to the jet launch region within the SKA telescopes sensitivity and resolution.\\
Beyond individual sources, SKAO will examine large samples of YSO jets across the Galaxy, enabling statistical studies of jet velocities as a function of mass, age, and environment. Its data will synergise with ALMA mapping of molecular outflow, JWST spectroscopy, and Gaia distances. 

\section{Molecular complexity in YSO outflows revealed by shocks} \label{sec:SKA_SC5_molecules}
As illustrated in Sect.~\ref{sec:intro}, shocks occurring along protostellar outflows have a stratified density and temperature structure, and the characterisation of their chemistry requires multi-wavelength observations. While sub-millimeter observations trace complex O-bearing molecules, the study of carbon-chain chemistry in UV-irradiated outflow cavities requires observations at longer wavelengths. Large carbon chains have their rotational transitions peaking at cm-wavelengths at low temperatures, making the mm range insufficient to capture their full emission. However, current cm-wavelength facilities do not yet offer the angular resolution and sensitivity required to detect and spatially resolve these tracers. The capabilities of the SKA telescopes will be crucial to explore complex carbon chain chemistry. By coupling observations of carbon chains in the cm range with O-bearing species in the (sub-)mm range, we will gain key insights into the chemical stratification and the physical structure of the outflow cavity, including the role of UV irradiation and shock processing across different layers. These observational diagnostics are particularly powerful when interpreted alongside cutting-edge 3D-MHD simulations of star-forming regions, which now include significantly improved microphysical modelling. Recent advances in numerical simulations - both in terms of dynamics (e.g., wind and jet launching, radiation processes; see \citealt{Grudic21}) and chemistry/microphysics (e.g., complex molecules, isomers, isotopologues; see \citealt{Bovino20, Lupi21}) - have made it possible to generate highly detailed synthetic observations. When combined with data acquired with the SKA telescopes, this wealth of simulated information will provide crucial constraints on the evolution of chemical complexity and the role of dust physics (see Sect.~\ref{sec:SKA_SC6_dust growth}) during the earliest stages of star and planet formation.

\subsection{Cyanopolyynes as probes of outflow shocks}\label{subsec:astrochemisty}
\begin{figure}
    \begin{center}
    \includegraphics[width=0.95\textwidth]{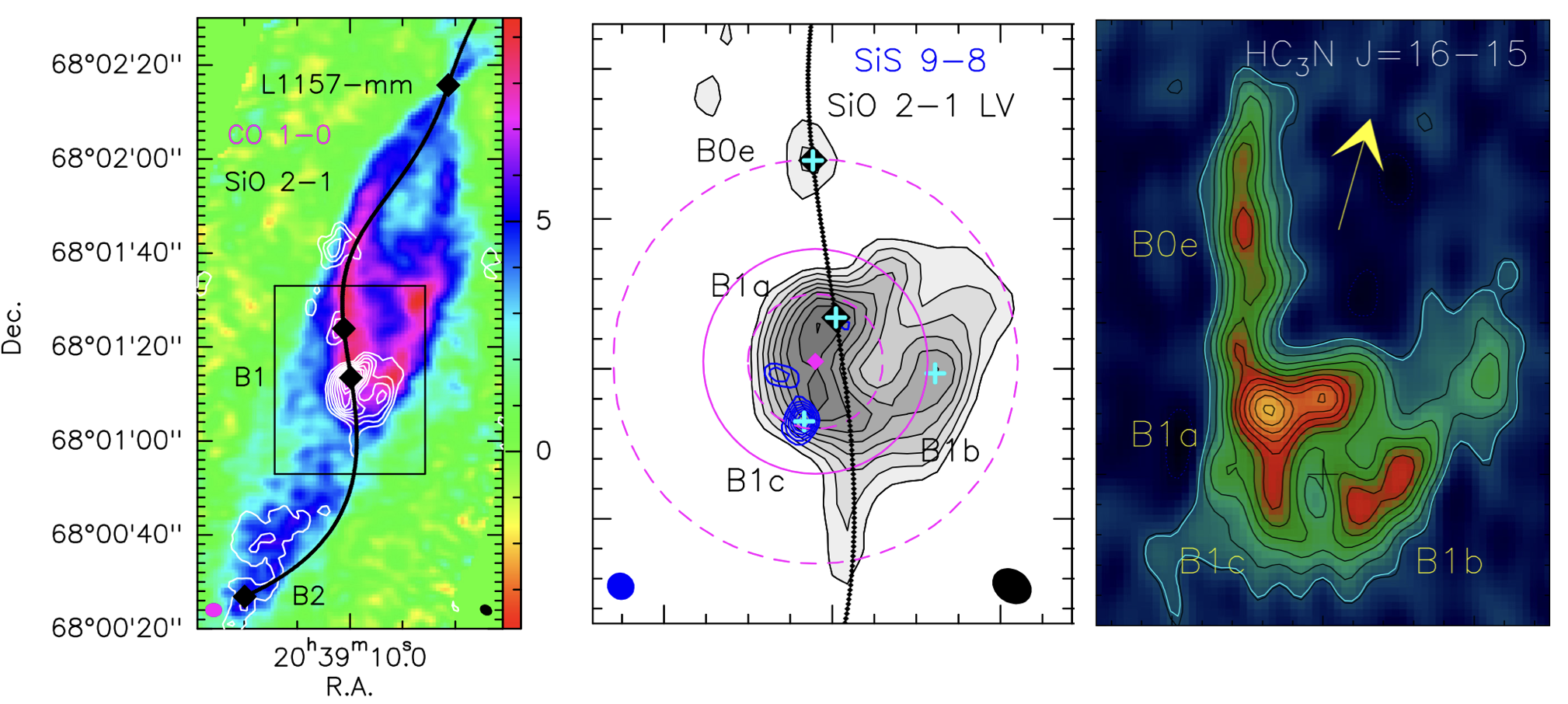}
    \caption{({\it Left}) The protostellar shock L1157-B1 along the southern lobe of the outflow driven by L1157-mm. (Middle/right): Zoom-in around L1157-B1, where the abundance of several molecules is mapped; Metal-bearing molecules (e.g. SiO and SiS; {\it middle}) probe grain core sputtering; ({\it right}) HC$_3$N is released from the grain mantles. Adapted from \citet{Podio2017} and \citet{Benedettini2013}.}
    \label{fig:l1157_hc3n_sio-sis}
    \end{center}
\end{figure}
\begin{SCfigure}
    \centering
    \includegraphics[width=0.62\textwidth]{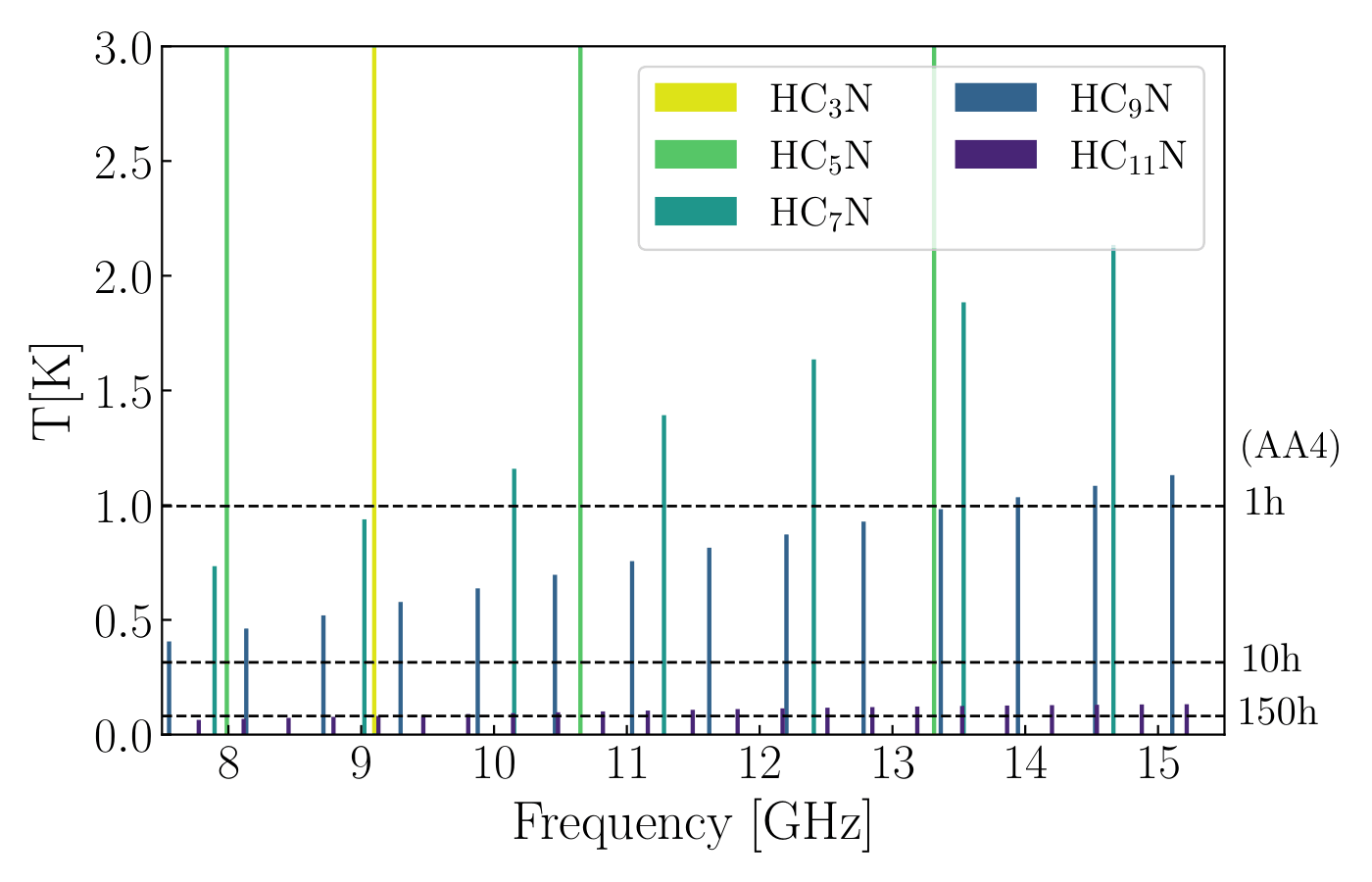}
    \caption{Predicted (LTE) line intensities of cyanopolyynes in high-mass protostellar outflows, obtained for $T=20$~K, FWHM = 4~km\,s$^{-1}$, and a filling factor of 1. We assumed $N$(HC$_5$N)$ = $3$\times$10$^{15}$~cm$^2$ (\citealt{Hoque2025}; HC$_3$N/HC$_5$N = 3), following \cite{Bianchi2023b} for the other cyanopolyynes. The horizontal dashed lines indicate the 3$\sigma$ predictions of the SKAO sensitivity calculator at 12.5~GHz with the SKA-Mid AA4 (including MeerKAT antennas; see text).}\label{fig:predictions-HCnN-CnH}
\end{SCfigure}

As detailed in Sect.~\ref{sec:mm-telescopes}, the abundance of typical outflow tracer species is enhanced by dust sputtering and shattering in shocks. These processes cause the release of the molecules from grain mantles and cores and activate the warm gas-phase chemistry.
As first noted by \citet{Bachiller1997} also Cyanoacetylene (HC$_3$N) is a powerful probe of shocked gas, with an increase of its abundance of $\sim$100 in the prototypical protostellar shock L1157-B1 (see Fig. \ref{fig:l1157_hc3n_sio-sis}).
Follow-up observations by \citet{Mendoza2018} show emission of several HC$_3$N isotopologues, as well as HC$_5$N in L1157-B1. Moreover, recent surveys at mm wavelengths have shown that HC$_3$N is an optimal tracer of outflowing gas also in high-mass star forming regions \citep[HMSFRs; e.g., ][]{Taniguchi2018,Lu2021,Wang2022}, with the detection of blue- and red-shifted outflowing gas in about 1/3 of the massive star-forming regions surveyed by the ATOMS programmes \citep{Hoque2025}. Longer cyanopolyynes, such as HC$_5$N and HC$_7$N, have also been detected in HMSFRs \citep[e.g., ][]{Taniguchi2018,Wang2022}.
The typical column density ($N$) of HC$_3$N is a few $10^{13}$ cm$^{-2}$ (at $T=20-50$ K) in the shock driven by the low-mass protostar L1157-mm \citep{Benedettini2013,Mendoza2018}, while $N$(HC$_3$N) ranges from a few $10^{14}$ cm$^{-2}$ to $10^{16}$ cm$^{-2}$ towards the outflows driven by massive YSOs and observed at $1''-2''$ resolution.
These mm-observations demonstrated that HC$_3$N emission is strongly correlated with SiO, thus suggesting that HC$_3$N is a reliable  outflow/shock tracer. On the other hand, the emission peak of heavier cyanopolyynes (HC$_{2n+1}$N, with $n>2$) in the cold outflow cavities ($T \sim20$ K) is found at cm wavelengths. 
In this context, SKA-Mid Band 5a and 5b (4-15 GHz) observations open a new window to characterise complex carbon chain chemistry in shocks at high angular resolution and sensitivity.
Figure \ref{fig:predictions-HCnN-CnH} shows the detectability of cyanopolyynes along protostellar outflows with SKA-Mid Band~5b receivers. Adopting the high-mass case as reference, we assumed $N$(HC$_3$N)~=~10$^{16}$ cm$^{-2}$, and scaled the column densities of longer cyanopolyynes from \cite{Bianchi2023} and \cite{Giani25}. %\citet{Giani2025}. 
The SKA-Mid sensitivities reported in Fig.\ref{fig:predictions-HCnN-CnH} were derived using the following parameters: a central frequency of $\nu_0=12$~GHz (with a 5~GHz total bandwidth), a spectral resolution of 1.5~km~s$^{-1}$, and Briggs weighting with robust=0, optimised to achieve an angular resolution of $\sim$2''. Based on these sensitivities, the detection of cyanopolyynes up to HC$_9$N will be possible in $\lesssim$10~h in AA4 configuration (including MeerKAT antennas). This makes the observations related to such a science case feasible even during the SKAO Science Verification with AA$^*$ configuration to detect lighter cyanopolyynes (i.e. $\leq$ HC$_{7}$N). The detection of heavier species (e.g. HC$_{11}$N) will require longer integrations, potentially hundreds of hours with the AA4 configuration. The expected $N$(HC$_3$N) in low-mass star-forming regions may drop remarkably compared to the high-mass regime (from $\sim$10$^{16}$~to a few 10$^{13}$~cm$^{-2}$), and similar results may require an integration time 10 to 50 times larger using SKA-Mid AA4 configuration.

\subsection{Astromineralogy as probe of dust-refractory cores}\label{subsec:astromineralogy}
Astromineralogy investigates minerals in space, providing insights into the constituents of planetary cores. The properties of dust grains are largely unknown, as the composition of solid-phase dust often eludes direct observation, and the typical IR bands are difficult to identify unambiguously (e.g. \citealt{Kimura05}). The volatile compounds on the dust mantles evaporate at relatively low temperatures ($\lesssim$~100~K) and can be observed through line emission at mm wavelengths.
On the other hand, the refractory cores of dust grains \citep{Hansley2021}, containing silicon and aluminium, pose significant challenges for observation due to their high sublimation temperatures (thousands of Kelvin). 
As reported in Sect.~\ref{sec:mm-telescopes}, shocks along protostellar jets are perfect laboratories to reveal the chemical enrichment of the gas phase due to the atoms and molecules lifting off the refractory component. Observing refractory species in protostellar shocks may be easier due to the fact that shocked areas have sizes of hundreds to thousands of au, which can be more easily resolved than the inner circumstellar region where the dust is sublimated.\\
The refractory cores are expected to contain Si-, and Al- molecular species, as well as salts such as e.g. NaCl and KCl. 
These species have been observed in the atmospheres of asymptotic giant branch (AGB) and post-AGB stars \citep[e.g.][and references therein]{Cernicharo1987,Sanchez2022}, and recently in the sublimation region of disks around massive protostellar sources, such as Orion SrcI 
\citep{Ginsburg2019a,Ginsburg2023b}.\\
On the other hand, Si-bearing species have been routinely detected in shocked regions along outflows driven by low- and high-mass protostars in the form of SiO \citep[e.g.][and references therein]{Podio2021}. The chemical rich protostellar shock L1557-B1 has been also detected in SiS \citep{Podio2017}, as well as of the Cl-bearing molecule  HCl \citep{Codella2012}. 
The SKA-Mid Band 5b frequency range (8–15 GHz), covers the fundamental rotational transitions ($J$ = 1–0, $v$ = 0) of NaCl, KCl, MgCl, AlCl, and CaS, as well as the fundamental transitions of the vibrationally excited states  of NaCl ($v$ = 1, 2, 3, 4), and KCl ($v$ = 1).
As shocks are characterised by a stratification of density and temperature conditions, down to $\sim$ 20 K \citep[CO,][]{Lefloch2012}, the extremely low-excitation ($E_{\rm u}$ $\sim$ 1 K) of the $J$ = 1–0, $v$ = 0 lines are expected to probe the low-temperature regime ($\sim$10~K), while the $J$ = 1–0, $v$ = 1, 2, 3, 4 transitions of NaCl and KCl, with upper energy levels ($E_{\rm u} =300-2000$ K) are expected to trace regions with temperatures exceeding 100 K. Astrochemical models of the formation of metal-bearing elements subsequent to the release of the grains cores in shocks are not available, therefore we lack 
estimates of the expected column densities of NaCl, KCl, MgCl, AlCl, and CaS in shocked regions. In this context, the SKAO may open a new window for the detection of metal-bearing molecules in protostellar outflows, and hence for the comprehension of the composition of dust cores and their reprocessing in shocks.\\
Finally, metal-sulfides have recently been detected toward the Giant Molecular Cloud G+0.693-0.027 located in the Galactic Centre \citep{Rey-Montejo24, Jimenez-Serra25}, whose chemistry is characterised by a large-scale shock, similar to those found in molecular outflows. The chemistry of refractory materials in outflows can then be compared to those found in molecular clouds of the Galactic Centre, which has also been proposed to be observed with the SKA-Mid (see Chapter by \citealt{Traficante01.2026.SKA} and \citealt{Schoedel01.2026.SKA}). These observations will provide key information about the universality of the refractory content of dust grains across different metallicity regimes in the Galaxy.

\section{Dust growth and transport along molecular outflows} \label{sec:SKA_SC6_dust growth}

\begin{figure}
\centering
\includegraphics[width=0.95\textwidth]{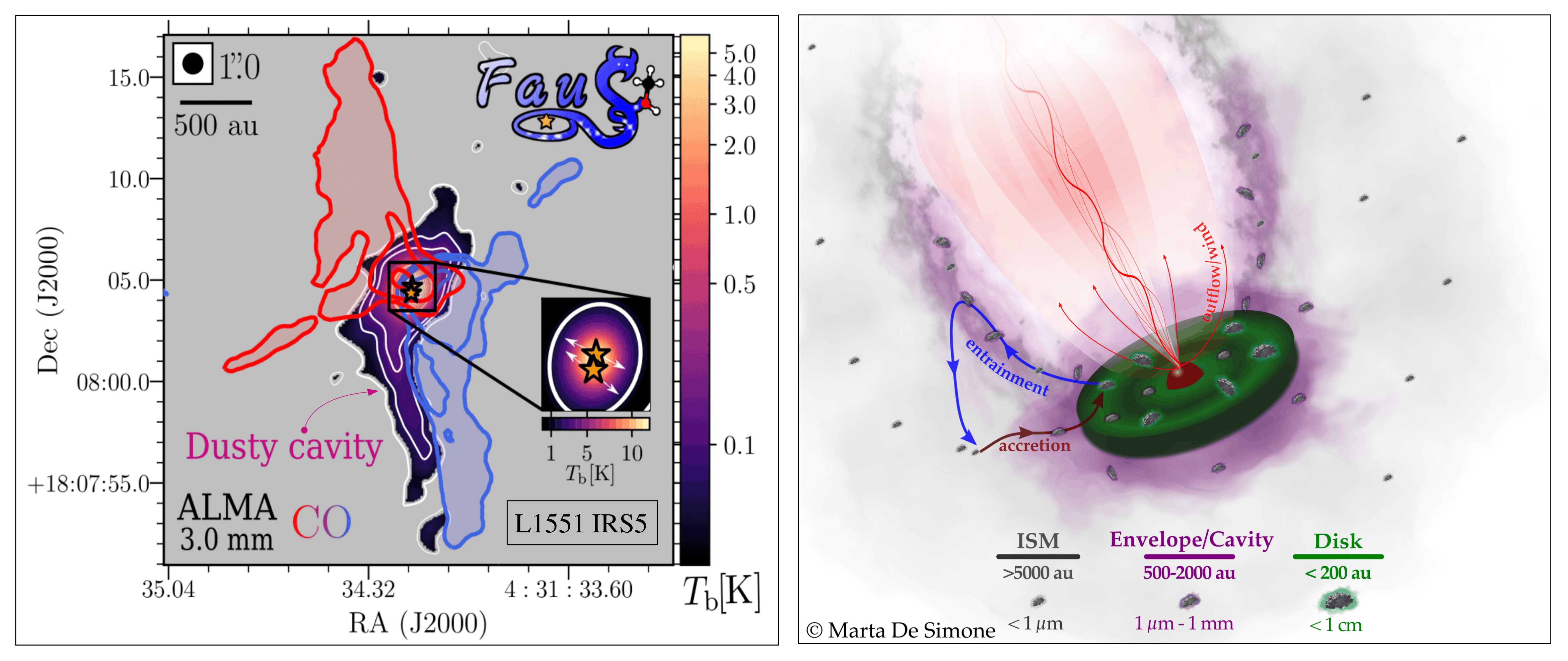}
  \caption{(Left) The L1551 IRS5 Class I binary system from \cite{Sabatini2025}. colour map: Thermal dust continuum emission at 3~mm (from the FAUST ALMA-LP) mapping the circumbinary disc and the dusty cavity walls structure. White contours mark the [3, 6, 10, 100]$\sigma$, with $\sigma= 0.08$~mJy~beam$^{-1}$. Transparent red and blue areas: The red- and blue-shifted CO~(2-1) outflow emission. (Right) Sketch illustrating how dust grains could be lifted from the young discs into the surrounding envelope by the outflow activity.} 
  \label{fig:dustycavities}
\end{figure}
As evidence grows that planets may form during the earliest phases of star formation \citep[e.g., ][]{Tychoniec2020, Ohashi23, Maureira2024, Maureira25}, it becomes crucial to determine when and how rapidly dust evolves into planetesimals. While the Chapter by \cite{Garufi01.2026.SKA} discusses dust growth in protoplanetary discs, here we focus on earlier stages, when dense protostellar envelopes funnel material onto forming discs and possibly interact dynamically with protostellar outflows. These envelopes, with visual extinctions from tens to hundreds, act as bridges transferring dust and angular momentum from parent cores to discs. Whether dust grows during collapse, and what properties characterise the dust delivered to the youngest discs, remains debated.\\
Multi-wavelength observations constrain the dust opacity spectral index ($\beta$), which depends on the grain size (\citealt{Testi2014, Ysard2019}). In the ISM, $\beta \sim 1.7$ corresponds to sub-micron grains, while $\beta \lesssim 1$ in protoplanetary discs signals millimeter-sized grains (\citealt{Tazzari2021, Garufi2025}). Envelopes, however, are difficult to study due to the challenges in separating disc and envelope emission. Early {\it Submillimeter Array} (SMA), {\it Combined Array for Research in Millimeter-wave Astronomy} (CARMA) and ATCA studies (\citealt{Kwon2009, Miotello2014}) hinted at dust growth but suffered from large uncertainties. Later works such as \citet{Bracco2017} found $\beta$ ranging from $\sim$0 in Class II discs to $\sim$1.7 in Class 0 protostars, though others found no deviation from the ISM value (\citealt{Agurto-Gangas2019}). Recent observational campaigns, such as the CALYPSO (IRAM-PdBI) and the ``Fifty AU STudy of the chemistry in the disk/envelope system of Solar-like protostars'' \citep[ALMA-FAUST, ID:~2018.1.01205.L, PI: S. Yamamoto; see][]{Codella2021} Large Programmes, measured $\beta$ in $\sim$20 protostars (\citealt{Galametz2019, Cacciapuoti2025}). About 50\% show evidence for large ($>$100~$\mu$m) grains, leaving open whether dust growth is intrinsic or observationally biased. Simulations struggle to reproduce such rapid growth \citep[e.g.][]{Ormel2009, Bate2022}, suggesting that grains may instead form in discs and be entrained by outflows (\citealt{Tsukamoto2021, Bhandare2024, Uchimura2025}). Observations indeed show correlations between low $\beta$ values and stronger outflows (\citealt{Cacciapuoti2024}), polarised emission along cavity walls (\citealt{Hull2020, LeGouellec2023}), and low $\beta$ in outflow cavities (\citealt{Sabatini2024, Sabatini2025}; see Fig.~\ref{fig:dustycavities}), all hinting at grain transport. Such recycling could affect growth timescales, outflow heating, and compositional mixing, similarly to Solar System evidence for inward-formed calcium-aluminium-rich inclusions (CAIs) found in outer asteroid families (\citealt{Morbidelli2024}).\\
However, $\beta$-measurements are prone to contamination from free-free, gyrosynchrotron, and anomalous microwave emission (\citealt{Dickinson2018}). At 3~mm, for example, up to 20\% of flux can arise from free-free processes (\citealt{Galametz2019, Hull2016}). These components vary over time, requiring multi-epoch, low-frequency ($<90$ GHz) data to disentangle them properly. The SKA-Mid, together with multi-band ALMA observations, will provide the resolution, sensitivity, and frequency coverage needed to build high S/N, resolved SEDs from sub-mm to cm wavelengths, allowing robust dust evolution studies from envelopes to discs. Taking the example reported inFig.~\ref{fig:dustycavities} for the L1551 IRS5 Class I binary system, we can estimate the sensitivity required for observing the dusty cavity wall with SKA-Mid Band-5b. In this system, an average dust opacity spectral index $\beta$$\sim$0.5 was measured along the entire (<2000~au) extent of the dusty cavity wall detected with ALMA Band 3 ($\sim$100~GHz). Under the Rayleigh-Jeans approximation, and assuming the dust emission follows a power law, $F_\nu \propto \nu^{\alpha_{\rm mm}}$ \cite[e.g., ][]{Testi2014}, with a spectral index $\alpha_{\rm mm}\sim 2.5$ \citep{Sabatini2025}, we estimate that a sensitivity of 0.4~$\mu$Jy/beam at 13 GHz is needed to ensure a  $>$7$\sigma$ detection at the edge of the cavity wall. This sensitivity is achieved by SKA-Mid AA4 (including the MeerKAT antennas) in less than 13~h of integration time, making this science case suitable for the SKAO Cycle 0. These calculations assume a continuum bandwidth of 5 GHz centered at 12 GHz, resulting in a resolution of $<$1$''$ (Briggs weighting, robust=1), which is sufficient to spatially resolve the cavity walls. 

\section{Conclusions and final remarks}
In this Chapter, we have discussed how the SKA Observatory will revolutionise our understanding of outflows and jets associated with YSOs across the full range of protostellar masses. Leveraging its maximum baseline length ($\sim$150~km for SKA-Mid\footref{webSKAmid} and $\sim$74~km for SKA-Low\footref{webSKAlow}) as well as its unprecedented collecting area ($3.3\times10^4$~m$^2$ and $4\times10^5$~m$^2$ for SKA-Mid and SKA-Low, respectively)\footref{webSKAmid}$^,$\footref{webSKAlow}, the SKA telescopes are unique facilities capable of bridging the crucial gap between the observed properties of YSO-associated jets and outflows and theoretical expectations.
Crucially, this transformative investigation is not confined to the distant future, as almost all the scientific cases presented here are feasible early in the SKAO's operational timeline, beginning with Science Verification. For example, the sensitive detection of RRLs with SKA-Mid will be possible by exploiting the initial SKA-Mid~AA$^*$ configuration even in relatively distant ($d>1$~kpc) YSOs. Similarly, the search for small cyanopolyynes ($\text{HC}_5\text{N and}~\text{HC}_7\text{N}$) in cold outflow cavities and detailed study of the B-field via polarisation studies in protostellar jets, beyond the single case identified so far, are expected to be key achievements of SKA-Mid in its initial configuration. \\
Nevertheless, two main scientific cases require the enhanced SKA-Mid performance provided by the AA4 configuration. Firstly, studying the launch mechanism in the vicinity of YSOs will require physical resolutions of a few au around the YSO. This will only be possible by exploiting the maximum extension of the SKA-Mid baselines, which will likely become available starting from SKAO Cycle 1, providing an exceptional resolution of approximately 4 au in sources at a distance of 200 pc (see Sect.~\ref{subsec:SKAmid_launching}). Secondly, the study of emission associated with heavier cyanopolyynes, although possible during Science Verification, will require significantly shorter integration times with the AA4 configuration, including in forming protosolar analogues (see Sect.~\ref{subsec:astrochemisty}).\\
In this context, a crucial future requirement involves the development of the SKA-Mid Band~6 receiver (15-50 GHz), representing a key capability upgrade within a future SKAO Development Programme\footnote{\url{https://www.skao.int/en/science-users/118/ska-telescope-specifications}}. Such a receiver would be essential for removing the critical frequency gap that currently exists between 15~GHz (the upper limit of SKA-Mid Band 5b receiver) and 35~GHz (the lower limit of the ALMA Band 1 receiver). Opening this window will be essential for reliably characterising the continuum emission along the jets and associated outflow cavity walls, and for accessing a frequency regime in which many iCOMs have brighter transitions than those found in the SKA-Mid Band~5b\footnote{Further details are available  consulting the SKA Memo 20-01, accessible at \url{https://www.dropbox.com/scl/fi/zi4pwvb3noa66htybsrjo/20-01-Beyond-Band-5.pdf?rlkey=ocoy6rjwl0k0eira8p02fgq22&e=1&dl=0}}.

\section*{Acknowledgments}
\footnotesize{\it We thank the anonymous referee for the constructive and insightful comments. G.S., Cl.Co. and L.P. acknowledge financial support under the National Recovery and Resilience Plan (NRRP), Mission 4, Component 2, Investment 1.1, Call for tender No. 104 published on 2.2.2022 by the Italian Ministry of University and Research (MUR), funded by the European Union – NextGenerationEU-Project Title 2022JC2Y93 Chemical Origins: linking the fossil composition of the Solar System with the chemistry of protoplanetary disks – CUP J53D23001600006 – Grant Assignment Decree No. 962 adopted on 30.06.2023 by the Italian Ministry of Ministry of University and Research (MUR); the project ASI-Astrobiologia 2023 MIGLIORA (``Modeling Chemical Complexity'', F83C23000800005); the INAF-GO 2024 fundings ICES, the INAF-GO 2023 fundings PROTOSKA (``Exploiting ALMA data to study planet forming disks: preparing the advent of SKA'', C13C23000770005); the INAF Mini-Grant 2022 “Chemical Origins” (PI: L. Podio) and the INAF Minigrant 2023 TRIESTE (``TRacing the chemIcal hEritage of our originS: from proTostars to planEts''; PI: G. Sabatini). G.B., J.M.G., E.D.M., and A.S.-M.\ acknowledge support from grant PID2023-146675NB-I00 (MCI-AEI-FEDER, UE). G.B. acknowledges financial support from grant CEX2024-001451-M funded by MICIU/AEI/10.13039/501100011033. C.C.-G. acknowledges support from UNAM DGAPA-PAPIIT grant IG101224. J.G.M., E.D.M. and A.S.-M.\ also acknowledge support from the programme Unidad de Excelencia María de Maeztu CEX2020-001058-M; and from the RyC2021-032892-I grant funded by MCIN/AEI/10.13039/501100011033 and by the European Union `Next GenerationEU'/PRTR. A.M.-H. and I.J.-.S acknowledge funding from grant PID2022-136814NB-I00 funded by the Spanish Ministry of Science, Innovation and Universities/State Agency of Research MICIU/AEI/ 10.13039/501100011033 and by “ERDF/EU”. A.M.-H. has received support from grant MDM-2017-0737 Unidad de Excelencia ``María de Maeztu'' Centro de Astrobiología (CAB, CSIC-INTA) funded by MCIN/AEI/10.13039/501100011033. G.A., M.O. and G.B.-C. acknowledge financial support from grants PID2023-146295NB-I00 and CEX2021-001131-S, funded by MCIN/AEI/10.13039/501100011033. E.B. acknowledges the support from the Italian Ministry for Universities and Research under the Italian Science Fund (FIS 2 Call - Ministerial Decree No. 1236 of 1 August 2023) and the Next Generation EU funds within the National Recovery and Resilience Plan (PNRR), Mission 4 - Education and Research, Component 2 - From Research to Business (M4C2), Investment Line 3.1 - Strengthening and creation of Research Infrastructures, Project IR0000034 – ``STILES - Strengthening the Italian Leadership in ELT and SKA''. G.B.-C. acknowledges support from grant PRE2018-086111, funded by MCIN/AEI/ 10.13039/501100011033 and by `ESF Investing in your future. S.B. acknowledges BASAL Centro de Astrofisica y Tecnologias Afines (CATA), project number AFB-17002. M.P. acknowledges the INAF grant 2023 MERCATOR (``MultiwavelEngth signatuRes of Cosmic rAys in sTar-fOrming Regions'') and the INAF grant 2024 ENERGIA (``ExploriNg low-Energy cosmic Rays throuGh theoretical InvestigAtions at INAF''). S.B. acknowledges BASAL Centro de Astrofisica y Tecnologias Afines (CATA), project number AFB-17002. A.T. acknowledges funding from the European Research Council in the ERC synergy grant ``ECOGAL – Understanding our Galactic ecosystem: From the disk of the Milky Way to the formation sites of stars and planets'' (project ID 855130).}

\bibliographystyle{abbrvnat-maxbibnames4}
\bibliography{chapter} 
\end{document}